\documentclass[iop]{emulateapj}   
\usepackage{amssymb, amsmath, amsfonts}
\usepackage{graphicx, shadow}
\usepackage{epstopdf}
\usepackage{rotating}
\usepackage{wasysym}
\usepackage{verbatim}
\usepackage{natbib}

\begin{document}

\title{K+A Galaxies as the Aftermath of Gas-Rich Mergers: Simulating the Evolution of Galaxies as Seen by Spectroscopic Surveys}
\shorttitle{Poststarbursts from Mergers}
\shortauthors{G. Snyder et al.}


\author{Gregory F Snyder\altaffilmark{1}, Thomas J Cox\altaffilmark{2}, Christopher C Hayward\altaffilmark{1}, Lars Hernquist\altaffilmark{1}, Patrik Jonsson\altaffilmark{1}}

\altaffiltext{1}{Harvard-Smithsonian Center for Astrophysics}
\altaffiltext{2}{Carnegie Observatories}

\journalinfo{ \apj Accepted Version, August 8, 2011}
\slugcomment{ \apj Accepted Version, August 8, 2011}
\keywords{galaxies:evolution, galaxies:interactions, galaxies:starburst, methods:numerical}


\begin{abstract}
Models of poststarburst (or ``K+A'') galaxies are constructed by combining fully three-dimensional hydrodynamic simulations of galaxy mergers with radiative transfer calculations of dust attenuation. Spectral line catalogs are generated automatically from moderate-resolution optical spectra calculated as a function of merger progress in each of a large suite of simulations.  The mass, gas fraction, orbital parameters, and mass ratio of the merging galaxies are varied systematically, showing that the lifetime and properties of the K+A phase are strong functions of the merger scenario.  K+A durations are generally $\lesssim$ 0.1-0.3 Gyr, significantly shorter than the commonly assumed 1 Gyr, which is obtained only in rare cases, owing to a wide variation in star formation histories resulting from different orbital and progenitor configurations.  Combined with empirical merger rates, the model lifetimes predict rapidly-rising K+A fractions as a function of redshift that are consistent with results of large spectroscopic surveys, resolving tension between the observed K+A abundance and that predicted when one assumes the K+A duration is the lifetime of A stars ($\sim$1 Gyr).  These simulated spectra are spatially resolved on scales of about 1 kpc, and indicate that a centrally-concentrated starburst causes the Balmer absorption strengths to increase towards the central few kiloparsecs of the remnant.  The effects of dust attenuation, viewing angle, and aperture bias on our models are analyzed.  In some cases, the K+A features are longer-lived and more pronounced when AGN feedback removes dust from the center, uncovering the young stars formed during the burst.  In this picture, the K+A phase begins during or shortly after the bright starburst/AGN phase in violent mergers, and thus offers a unique opportunity to study the effects of quasar and star formation feedback on the gas reservoir and evolution of the remnant.  Analytic fitting formulae are provided for the estimates of K+A incidence as a function of merger scenario.
\end{abstract}


\section{Introduction} \label{s:intro}

Generally speaking, bright galaxies fall into two broad visible-spectrum categories: star-forming, blue spiral galaxies, and passively-evolving, red elliptical galaxies \citep{hubble26}.  Two long-standing questions for studying galaxy formation are:  How does each type of galaxy form?  Do galaxies transform from one type to another, and if so, how?  

Early studies of the relationship between these classes as a function of lookback time focused on galaxy clusters.  \citet{oemler74} and \citet{bo78} concluded that intermediate-redshift ($z \sim 0.5$) clusters have a much larger fraction of blue, star-forming galaxies than nearby clusters: the ``Butcher-Oemler effect''.  \citet{dressler83} found that in one cluster, these blue cluster members were not ``normal'' spiral galaxies as we might see in the local Universe, but had spectra indicating either nuclear activity (e.g. Seyfert) or a recently quenched star-forming episode.  The latter, so-called ``E+A" galaxies, named as the combination of an early-type galaxy spectrum (E) with an A type stellar spectrum, were fitted best by an old stellar population plus a starburst component whose development ended within the last $\sim1$ Gyr; they have strong Balmer absorption features but weak or undetected nebular emission lines.  Also known as ``K+A'' (for K type stars) or ``poststarburst'' galaxies, E+A galaxies were seen as occupying an important transitional state between the star-forming and passive populations.  These objects are commonly selected from spectroscopic surveys as having H$\delta$ equivalent width greater than some value (typically roughly 5\AA) in absorption and [O II] emission equivalent width less than roughly 2\AA.  The latter limit corresponds to a cut in specific star formation rate at roughly $10^{-11}$yr$^{-1}$, or a star formation rate below about 2 M$_{\odot}$yr$^{-1}$ for the most massive galaxies we consider in this work \citep{kennicutt98}.

Numerous studies of K+A galaxies have been undertaken to explore this paradigm of galaxy evolution.  K+A galaxies were found to be a significant fraction ($\sim 0.2$) of rich cluster members at $0.2 \lesssim z \lesssim 0.6$ \citep{couch87, dressler90, fabricant91, belloni95, balogh99, poggianti99}, but are fewer than 1\% of cluster members at $z < 0.1$ \citep{dressler87}.  Thus, a cluster-specific mechanism was thought to be responsible for quenching the star formation in these galaxies at intermediate redshifts, leaving fewer potential progenitors for the present time.  Unbiased spectroscopic surveys revealed that although K+As seem to be a sizable fraction of cluster galaxies, they also exist in substantial numbers in lower density regions \citep{zabludoff96, blake04, goto07, yan08, vergani09} out to $z \sim 1$.  In fact, the majority of K+A galaxies in the local universe do not reside in clusters \citep{quintero04}.  Thus, a cluster-specific mechanism cannot be the only way to make a K+A galaxy \citep[also see][]{yan08}, and a systematic study of alternative mechanisms is warranted.  

Detailed studies of the morphology, kinematics, and metal abundances of nearby poststarburst galaxies (primarily in the field) indicate that many of them are likely progenitors of bulge-dominated galaxies \citep{norton01, blake04, quintero04, goto07metals, yang08}.  Thus, studying K+A galaxies may give clues to the evolution of galaxies from star-forming disks into quiescent ellipticals in a variety of environments at different cosmological times.  \citet{lavery88} proposed galaxy-galaxy interactions as a mechanism for driving K+A formation.  A number of previously-mentioned studies of nearby K+As point out that a substantial fraction exhibit tidal features \citep{zabludoff96, chang01} or companion galaxies \citep{yamauchi08}, suggesting that galaxy interactions or mergers indeed play a role in K+A formation.  In addition, hydrodynamic simulations \citep[with simple star-formation laws, e.g.][]{mh94a,mh96} of gas-rich galaxy mergers can lead to global star formation histories (SFH) characterized by strong, short bursts.   Using simple stellar population synthesis (SPS), these bursts can lead to spectra with the lack of nebular star-formation indicators and strong Balmer absorption and that are found in poststarburst galaxies.  

These studies and others led to proposed K+A formation mechanisms that fall broadly into two classes:  (1) interaction of a star-forming disk with another galaxy, and (2) processes specific to massive clusters and halos.  While a non-cluster-specific mechanism is required for K+As in the field, the relative inefficiency of major mergers in systems of galaxies with high relative velocities implies that galaxy-galaxy interactions may not currently be sufficient for creating the large number of K+As in the highest-mass halos.  

Ram-pressure stripping \citep{gunngott72}, the removal of gas from disks traveling at high-speeds through a hot intracluster medium (ICM), is a likely candidate for reducing or ending the star formation in galaxies that fall into clusters.  Although this process alone may not be violent enough to trigger a K+A phase, numerical work indicates that such galaxy-cluster interactions may lead to compression of the disk gas, offsets of the disk gas with respect to the galaxy's central potential, or infall of stripped gas that remained bound to the subhalo \citep[e.g. see][]{schulz01, vollmer01, vangorkom04}.  These may lead to enhancements of the star formation rate, and thus a better chance of seeing strong Balmer lines in a galaxy's spectrum \citep{dressler83}.  However, these systems cannot dominate the population of bright K+A galaxies, because the vast majority of galaxies at these masses ($\gtrsim 10^9 M_{\odot}$) are centrals \citep[e.g.][]{yang_x09}, so they undergo their transition from star-forming to passive as central galaxies.  

In addition, for galaxies in halos above $\sim 10^{12} M_{\odot}$, cosmological accretion may lead to virial shocks and heating \citep{rees77} that shut down star formation.  Numerical simulations of this phenomenon \citep[e.g.][]{keres05, dekel06, birnboim07} indicate that galaxies in massive halos may experience a powerful burst-quench cycle leading to a post-starburst phase.  However, this mode is likely not a dominant contributor to the K+A population for several reasons.  In particular, gas will continue to cool and rain down on the galaxy \citep{keres_hernquist09}, so it is not certain that a complete (or fast enough) cessation of star formation will occur.  In addition, it is not clear if this mode can drastically alter galaxy morphology in a manner consistent with field K+A galaxies.  

In this work, we focus on K+A formation via galaxy-galaxy interactions.   Mergers have long been thought to be responsible for destroying stellar disks and turning them into dispersion-dominated ellipticals: the ``merger hypothesis'' \citep{toomre77}.  However, recent advances in numerical simulation methods have allowed the study of galaxy evolution via mergers in a significantly more quantitative way.  For example, it has recently become possible to ``age-date'' the time since the starburst in K+A galaxies \citep[e.g.][]{yang08}, and merger simulations can be used to test these methods by including a more sophisticated treatment of dust attenuation.  

In particular, using physical models also allows the study of realistic SFH outcomes for a given set of mergers.  Empirical models used in studies of K+A galaxies have often assumed very simple forms for the SFH induced by a merger.  These SFHs imply ``K+A lifetimes'' of order the lifetime of an A type star, 0.5-1.5 Gyr.  While it is true that for a given galaxy, the merger-induced burst can mimic this trivial single-aged stellar population (SSP), it is not clear if this is true for the ``typical'' merger involving a given disk.  In fact, with fixed progenitors, variations in the orbital parameters of a merger can lead to an enormous variety of dynamical and SFH outcomes \citep{barnes92, bh96}.  The merger or event rate inferred by surveys of K+A galaxies are sensitive to this distribution of the K+A duration timescale.  In the present work, we quantify this variation on the inferred K+A lifetimes of simulated mergers.  

In addition, the mechanism for the rapid shutoff of star formation is not known precisely.  Feedback from star formation and active galactic nuclei (AGN) activity, presumably leading to the expulsion of the remaining gas reservoir, has been inferred indirectly in a wide variety of observations \citep{kaviraj07, tremonti07, brown09}.  By varying the feedback prescriptions and strengths in the numerical models, simulations can test a large variety of these processes in a systematic way.  More generally, by studying poststarburst galaxies in the context of the extragalactic zoo and standard cosmology, we hope to study their implications for the role of mergers in driving galaxy evolution.  

In this paper, we define and study a set of K+A galaxy models which are the remnants of simulated galaxy mergers.  We have seen that the characteristic rapid inflow and consumption of gas during a major gas-rich merger is likely to be the cause of some K+A galaxies, and a recent study by \citet{wild09} used simulations from \citet{johansson09} to perform a similar comparison of the models to poststarburst data; the reader is encouraged to see those papers for another detailed investigation of the characteristics of poststarburst galaxies in merger simulations.  Our conclusions are broadly consistent with those reported in \citet{wild09}.  

The present work applies different spectrum synthesis and selection methods to a broad set of simulations by \citet{cox06_kinematics} to systematically quantify the variation of K+A properties with different merger parameters.  In particular, we utilize three-dimensional radiative transfer to study the post-starburst stellar populations using standard line strength indicators.  This procedure allows us to arbitrarily vary observing parameters such as the viewing angle and spectrograph fiber size and realistically measure the line strengths for a wide range of simulated mergers that include processes such as star formation, feedback, accretion onto a central supermassive black hole, and dust attenuation.

The primary goal of this work is to reconcile observed K+A numbers with realistic merger models and well-motivated merger rates.  In \S\ref{s:methods}, we outline in detail our methodology for generating spectral line catalogs from numerical simulations.  Readers wishing to bypass these details can skip to the relevant parts of \S\ref{s:basic_studies}, \S\ref{s:params}, \S\ref{ss:summaries}, or \S\ref{s:counts}.  In \S\ref{s:results}, we study the basic ramifications of our methods, highlighting the resulting variations owing to viewing angle, aperture bias, the presence of a bulge, and AGN feedback.  In \S\ref{s:params}, we systematically calculate the lifetime of the K+A phase as a function of the input merger parameters, and summarize the results in \S\ref{ss:summaries}, providing fitting formulae that can be used in semi-analytic modeling.  In \S\ref{s:counts}, we discuss the implications of our study by comparing observed K+A fractions to those inferred by our lifetime calculations.  We summarize and discuss the implications of these studies in \S\ref{s:discussion}, and conclude in \S\ref{s:conclusions}.  


\section{Methods} \label{s:methods}
We employ the {\sc Gadget 2} code \citep{springel05} to simulate galaxy merger scenarios.  In \S\ref{ss:sims} we describe the mergers studied in this work and the relevant physics included.  Stellar light and dust extinction are modeled post-simulation using the three-dimensional polychromatic radiative transfer code {\sc Sunrise} \citep*[][hereafter J06, J10]{jonsson06, jonsson09}, with specific models discussed in \S\ref{ss:sunrise}.  We produce spatially-resolved spectral energy distributions (SEDs) along multiple sight lines from 50x50 pixel cameras with each pixel spanning two physical kiloparsecs.  To select K+A galaxies, we mimic the data of optical spectroscopic surveys (\S\ref{ss:spectra}) and generate spectral line catalogs for each simulation. For this purpose, we use the following equivalent widths: the Balmer lines (H$\beta$, H$\gamma$, H$\delta$) = (4861\AA, 4341\AA, 4101\AA) \ and the [O II] $\lambda = $ 3727\AA \ transition (\S\ref{ss:catalogs}).  In Figure~\ref{fig:methodpanels} we demonstrate our aperture and spectrum fitting procedure.


\subsection{Galaxy Simulations} \label{ss:sims}

This study is one in a series based on numerical simulations of gas-rich galaxy mergers.  The calculations used here are essentially a subset of those employed by a number of other authors, for example \citet{cox06_kinematics}, \citet{hopkins05_qlife}, \citet{narayanan09_dog,narayanan10_smg}, \citet{robertson06}, \citet{wuyts09,wuyts09b,wuyts10}, \citet{younger08,younger09}, \citet{hoff09,hoff10}, and others.  The focus of this work differs in that we wish to generate and analyze mock spectral catalogs of merger remnants in order to quantify their observed poststarburst characteristics.  Hence, most of the discussion in this section will be on the application of the radiative transfer code {\sc Sunrise} (J06, J10) to the extraction of realistic spectra.  For a complete description of the physical models applied to the hydrodynamic methods in the context of this ongoing study, the reader is encouraged to see \citet*{sdh05} and \citet{cox06_kinematics}.  

{\sc Gadget 2} \citep{springel05} combines an accurate N-body gravity solver with a smoothed particle hydrodynamics (SPH) code that is fully conservative \citep{sh02} to perform cosmological or galaxy simulations.  Gravity dominates on galactic scales, but hydrodynamics is essential in many applications and serves as the basis for treating a great deal of challenging baryon physics.  Our simulations spawn star particles stochastically from the gas via a prescription designed to reproduce the Schmidt-Kennicutt law \citep{springel03}.  Each galaxy is seeded with a central massive black hole that undergoes Eddington-limited Bondi-Hoyle accretion.   These processes inevitably inject energy back into the surrounding gas (supernovae, jets, winds, radiation, etc), so our simulations model this by thermally coupling some of this energy to nearby gas particles \citep*{dsh05, springel05, sdh05}.  The feedback efficiency parameters used are identical to those used in previous studies \citep[e.g.][and subsequent works]{hopkins05_qlife, cox06_kinematics}: the radiative efficiency is 0.1, and the feedback energy couples dynamically to the surrounding gas with an efficiency of $\approx 5\%$. 

Our star-forming disk galaxies consist of dark matter halos, stellar and gaseous disks, and stellar bulges.  We place two disks on parabolic orbits with a variety of relative disk orientations.  Generally speaking, the major mergers (two galaxies equal in mass) proceed as follows.  As the two disks approach for the first time, strong non-axisymmetric forces cause the disks to form a transient bar, driving some material into the center \citep{bh91, bh96}.  How much material depends on the merger orientation as well as the properties of the original galaxy \citep{hopkins_disk09,hopkins_gas09}.  For example, \citet{mh94a, mh96} find that the presence of a stellar bulge can resist perturbations owing to this first passage.  

Ultimately, the torques produced during this close passage transfer angular momentum from the halos' mutual orbit into their constituents, leading to a direct coalescence of the two disks.  During this phase, the perturbing forces are so great that they completely disrupt the disks, sending much of the remaining gas into the center of the combined dark matter potential.  As the gas density increases in the central regions, star formation and central black hole growth proceed rapidly.  

In this study, we explore isolated binary mergers for which we vary galaxy mass, gas fraction, and bulge fraction, as well as their relative orientation and mass ratio.  


\subsection{Radiative Transfer} \label{ss:sunrise}

We generate mock optical observations of our galaxies in post-processing for all relevant times during each merger simulation.  Roughly speaking, knowledge of the global star formation rate and approximate wavelength-dependent dust attenuation is sufficient to produce a reasonable prediction of the integrated galaxy spectrum.  However, large spectroscopic surveys (in which poststarburst galaxies are identified) typically measure only a portion of the galaxy light in a spatially-biased way (fibers, slits), and we wish to explore how K+A selection varies with aperture size and viewing angle.  Therefore, we prefer a method that uses and preserves the information we get from the simulations about the locations of individual star particles and hence of distinct stellar populations, as well as the gas and dust.  This requires knowing the complete spatially-resolved star formation history (SFH) of the system, which we can follow via the histories of the stochastically-spawned SPH star particles.  We then must assign each particle a single-age stellar population (SPS) SED for a cluster of that mass, age, and metallicity.  Moreover, the wavelength-dependent attenuation cannot be reliably predicted from integrated quantities, as it depends on the geometry of stars, AGN, and dust, as well as dust grain composition and sizes, and scattering of light into the line of sight.  Thus we calculate the dust attenuation along each line of sight using knowledge of the surviving gas particles' spatial distribution and metallicity by assuming a dust-to-metals ratio.  

To these ends, we employ the {\sc Sunrise} polychromatic Monte Carlo radiative transfer code (described in J06 and J10).  This code performs all of the necessary steps above and allows us to conveniently vary important stellar modeling parameters for our study.  Below we discuss the principle components used to compile our final spectra.  In addition, we describe the modeling assumptions we make when limited by the finite resolution of the simulations.  These choices were made to be reasonable for local merger remnants, and were set before making any of the conclusions in \S\ref{s:results} or \S\ref{s:counts}.  In some cases (\S\ref{ss:systematics}), we varied these assumptions to explore their effect on a small subset of our simulations.  However, in this work we make no attempt to constrain or fit these modeling uncertainties, but rather treat them as a fixed set of inputs with which to test the overall picture.  

1.  \emph{Normal stars.}  To model stars of age $> 10^7$ years (10 Myr), we apply a high-optical-resolution ($\Delta \lambda/\lambda \approx 20,000$) {\sc Starburst99} SED \citep{leitherer99} with metallicity closest to that of the star particle.  We use the two-segment power-law initial mass function (IMF) of \citet{kroupa01}.  It is from these models that Balmer-series absorption enters our SEDs, since stellar clusters of a certain age will be dominated by A stars, originally defined by such strong absorption features.

2.  \emph{Young stars.}  For stars less than 10 Myr old, {\sc Sunrise} uses the SED templates of \citet{groves08}.  These models combine one-dimensional shock models for evolution of the H II and photodissociation regions (PDRs) around massive stars and clusters \citep{dopita05, groves08}, with the {\sc Mappings} dust and photoionization radiative transfer code \citep{binette82, sd93, groves04, allen08} to calculate the SED that emerges from the HII region and, optionally, the PDR, which is the region where the gas surrounding young stars transitions from being fully ionized to fully molecular.  These SEDs include nebular emission lines.  An uncertain tunable model parameter of this component is the time-averaged fraction, $f_{\rm PDR}$, of each H II region that we assume to be surrounded by the birth cloud from which the cluster was born (the PDR).  The dust in the PDR is a significant source of attenuation for young stars that are potentially the dominant contributor to the galaxy luminosity.  Much observational and theoretical work is attempting to constrain this parameter, and a number of discussions about its impact and use for studying such simulations can be found in J10, \citet{groves08}, \citet{narayanan09_smg}, and \citet{younger09}.  

Prior to running any simulations, we selected $f_{\rm PDR}=0.3$ as a realistic value for the majority of K+A galaxies at low redshifts.  This choice gives results that are identical to within 10\% (in line fluxes) to when the fiducial value of 0.2 from J10 is used.  These values of $f_{\rm PDR}$ correspond to molecular cloud clearing timescales of $\sim$1-2 Myr, which is consistent with observational estimates \citep[][and references therein]{groves08}.  As pointed out in J10, the {\sc MappingsIII} particles contribute weakly to the galaxy SEDs when the specific star formation rates are at or below those of typical local galaxies, which is what we expect for the majority of K+A galaxies found in local surveys.  See \S\ref{ss:systematics} for a discussion on the effect of $f_{\rm PDR} = 0.9$ on K+A selection.  

Another source of noise tied to the treatment of young stars owes to the stochasticity of the star formation prescription in the simulation.  Although it gives the expected overall star formation rates for each gas particle, the version of {\sc Gadget} we use for this study generates star particles using a finite table of random numbers for the Monte Carlo step.  This causes the formation of star particles at a given time to be more correlated than it ought to be, and the emission line star formation indicators reflect this by being highly variable from snapshot to snapshot even though on average the star formation histories are correct.  For all of our measurements, we smooth the spectral line catalogs over a 50 Myr box to limit the noisiness of the lightcurves owing to this effect.  This will add a comparable amount ($\sim 50$ Myr) of uncertainty to our measurements, but this is a small random effect.

3.  \emph{Dust attenuation}.  To attenuate the light originating from the source particles, {\sc Sunrise} performs Monte Carlo radiative transfer on an adaptively refined Cartesian grid.  For complete details, see J06 and J10.  We assume a dust-to-metals ratio of 0.4 \citep{dwek98} and Milky Way dust grain models from \citet{wd01} with \citet{draine07} updates.  Using the energy input into the dust by the source particles, {\sc Sunrise} computes self-consistently absorption, scattering, and subsequent infrared re-emission from dust.  The wavelength-dependent dust attenuation is calculated correctly from the assumed dust grain models, producing spectra for which the lines are attenuated realistically.  

The radiative transfer of dust attenuation by dense clumps requires subresolution considerations because the underlying hydrodynamic simulations do not have sufficient resolution to model the ISM on all scales.  Under the ISM treatment that we use \citep{springel03}, gas particles in {\sc Gadget} should be thought of as containing both a diffuse ``hot'' component and a clumpy ``cold'' component where the total masses of each are estimated, but we have no information about their physical distribution.  In this work we use the PDR model described above to treat the significant attenuation by the GMC from which a star forms, but assume that the covering of young stars by GMCs that are \emph{not} their birth clouds is negligible.  In this approach, we account for the cold phase (which can contain $\sim$ 90\% of the gas mass) only through the PDR model, which we use as the source spectrum for newly-spawned ($<$ 10 Myr) star particles.  This model is sometimes referred to as ``multi-phase ON''.  We believe this choice is appropriate for the conditions of most local K+A galaxies, where large reservoirs of obscuring dust are not available.  In \S\ref{ss:systematics} we briefly discuss an alternative choice. 

4.  \emph{Galaxy properties}.  We use [O II] as a star-formation rate indicator to select against poststarburst galaxies, so in order to capture the approximately correct metallicity dependence of this line, we assume the gas disks have an initial central stellar and gas-phase metallicity of 1.5$Z_{\odot}$ and gradient $-0.03$ dex\,kpc$^{-1}$, consistent with \citet{zaritsky94}.  We assume that the stars initially present in the disk were formed continuously over 14 Gyr and the bulge was formed in a single burst 14 Gyr ago.  

5. \emph{AGN}.  This implementation of {\sc Sunrise} uses the black hole accretion rate from the {\sc Gadget} simulation to estimate the bolometric luminosity of the central AGN, $L = \epsilon\dot{M}c^2$, where we use $\epsilon = 0.1$.  The code then inserts an unreddened luminosity-dependent broadband quasar SED from the templates in \citet{hrh07}.  This source is considered to be at the location of the central SMBH and its spectrum is treated in the same way as a star particle, meaning that it is subject to dust absorption and scattering.  While the underlying AGN templates used for this work include emission from narrow-line regions (NLR), the coarse interpolation used in this work to create the broadband AGN spectrum makes it impossible to measure this component to the lines (it is treated as continuum).  Thus the emission lines in our spectra solely owe to star formation, and are not contaminated by emission from NLRs.  


\subsection{Simulated Spectra} \label{ss:spectra}

For the results we present here on the lifetimes of nearby poststarburst galaxies, we generate mock catalogs designed to mimic Sloan Digital Sky Survey \citep{york00} data.  We fix the scale of our measurements by placing each galaxy at $z=0.1$.  The result of the {\sc Sunrise} procedure is, in general, a set of observations of the galaxy merger from several different camera angles.  We choose seven viewing angles distributed uniformly in solid angle.  One might think of each viewing angle for a given snapshot as a distinct measurement.  Thus our sample consists of $N_{\rm cameras} \times N_{\rm snapshots} \times N_{\rm simulations}$ measurements.  The mock data consist of the spectral flux density on a square grid of pixels for each camera.  For the present work, our pixels are at most 2 kpc on a side, which is sufficient to resolve typical spectroscopic apertures at $z \approx 0.1$.  

For a reasonable balance between spectral resolution and computation efficiency, we produce SEDs on an $R = \lambda/\Delta \lambda \approx 1000-1200$ wavelength grid for the range of interest, 3000-7000\AA.  However, owing to technical limitations at the time these calculations were made, we have used $R \approx 100$ for the {\sc Mappings} component of the galaxy SED (\S\ref{ss:sunrise}, component 2), oversampling it onto our R1200 grid.  Thus the nebular emission lines are wider than they should be given the spectral resolution of the output wavelength grid, but the total flux of each line is conserved.  This somewhat degrades the SEDs for which there is a substantial emission line component and renders impossible a two-component fitting of spectra in which there are comparably strong emission and absorption components.  To avoid confusion, we stick to simple line strength indicators and do not try to fit both emission and absorption components.  

We conserve the flux in the emission line component, so we still accurately measure the evolution of the total flux in the line during the ingress to the poststarburst phase as star formation/line emission turns off and A stars become dominant.  During the E+A phase and egress (as the A stars dominate then fade), the low-resolution {\sc Mappings} model is sub-dominant and our $\approx$ R1200 sampling becomes meaningful and effective for measuring the Balmer absorption lines as they shrink and disappear.  This means that we produce spectra (or sequences of spectra) in which the effective spectral resolution varies unrealistically in both time and wavelength.  The K+A selection criteria we employ will be sensitive to this, because ordinary Balmer absorption line strengths are calculated by measuring the equivalent width without any emission lines present, because they have been subtracted out.  However, we will apply an [O II] limit and use typically-observed metallicity profiles, so we expect any Balmer emission filling to be small for systems that we observe as having small [O II] emission fluxes.  We have checked this by running several simulations without the emission line component.  In these cases, our pipeline measures absorption equivalent widths that are the same to within $\sim 5\%$ for all snapshots that satisfy the [O II] criterion.  

\begin{figure*}
      \epsscale{1.0}
      \plotone{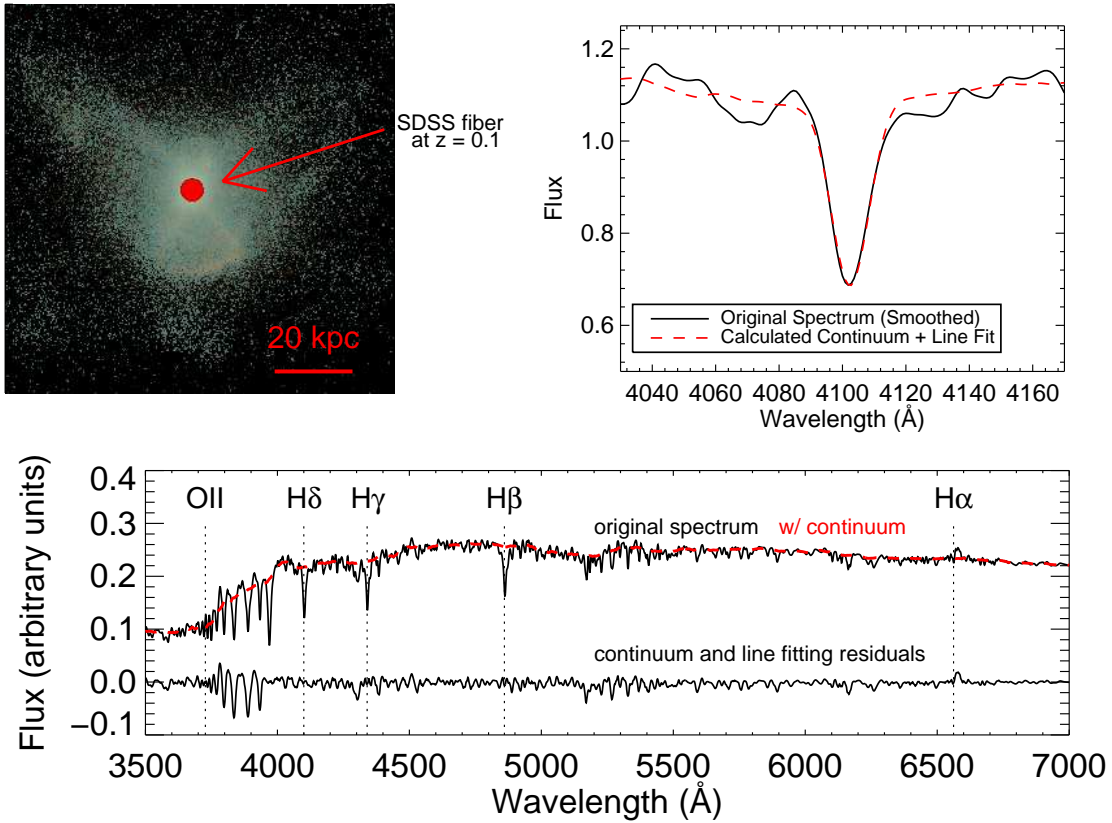} 
    
      \caption{An overview of our selection method.  Optical images are created for each of seven viewing angles for all simulation snapshots and scaled to z=0.1.  The light falling within a chosen aperture is taken to be the galaxy's spectrum in our default model.  A simple k-$\sigma$ rejection filter is applied to estimate the continuum level, and the lines of interest are then fitted with individual Gaussians.  The top left panel here shows the logarithm of the optical flux in g-r-i colors for one of our simulated K+A galaxies.  We overlay the SDSS spectrograph fiber scaled appropriately: this is what our default spectrum A seeks to mimic.  In the top right panel we show a zoom-in of our fit to the H$\delta$ line in the resulting spectrum.  In the bottom panel, we show the spectrum (black), our continuum fit (dashed red), and the residuals (black, arbitrary offset) after fitting for H$\epsilon$, H$\delta$, H$\gamma$, H$\beta$, H$\alpha$, and [O II].  This procedure allows us to create realistically-generated spectral line catalogs for a large number of simulations.  \label{fig:methodpanels} }
      \end{figure*}

We measure the simulated spectrum observed by each camera at every time in four ways.  By doing this for all of our simulations, we hope to get a rough idea of how these different observing modes can affect the selection of poststarburst systems formed from a variety of merger scenarios. 
\begin{description}
\item (Spectrum A)  Our default spectrum is measured by including the full radiative transfer model (elements 1-4 in the Section~\ref{ss:sunrise}) \emph{and by only using light that falls within a 3 arcsecond aperture}.  We center the aperture on the peak of the optical surface brightness via centroiding and weight the pixels according to their intersection with a 3 arcsecond diameter circle centered on this point.  For $z=0.1$, this corresponds to a diameter of approximately 6 kpc.  See Figure~\ref{fig:methodpanels} for a graphical description.
\item (Spectrum B)  We turn off element 3 of Section~\ref{ss:sunrise}, the attenuation by diffuse dust, and use the exact same aperture as for measurement A.  The attenuation in the Mappings subresolution model, item 2 in \S\ref{ss:sunrise}, is still included.  Note that we do not re-center the aperture, so cases A and B measure an identical line of sight.  
\item (Spectrum C)  We measure the integrated spectrum of the square camera, which is 100 kpc on a side.  We use this as a rough way to estimate the relative spatial concentration of the K+A component to the stellar populations.  
\item (Spectrum D)  We measure the integrated light as in C, but we turn off element 3 of Section~\ref{ss:sunrise}, the attenuation by diffuse dust.  
\end{description}

We point out that these methods generalize easily.  Our pixel-weighting scheme can make spectra using arbitrary spectrograph aperture shapes (slits, etc), and we scale the aperture consistently using the angular diameter distance to a given redshift and telescope properties.  Where needed, we assume a flat cosmology with $\Omega_m=0.3$ and $\Omega_{\Lambda}=0.7$.  In principle, it is feasible to mimic (in all but spectral resolution) any realistic observing mode.  

\subsection{Line Catalogs} \label{ss:catalogs}

Following a number of observational works, we use the rest-frame equivalent widths ($W_{line}$), and simple comparisons between them, as primary indicators of poststarburst activity.  This equivalent width is defined (ideally) as

\begin{equation} \label{eq:ew}
W_{\rm line} = \int_{\rm line}{\frac{I_{ \lambda} - C_{\lambda}}{C_{\lambda}} d\lambda},
\end{equation}
where $I_{\lambda}$ is the spectral flux in the line at wavelength $\lambda$, $C_{\lambda}$ is the spectral flux in the continuum at wavelength $\lambda$, and the integral is taken separately over each spectral line in the galaxy's rest frame.  

Thus, our goal is to generate catalogs of this quantity for each simulation snapshot, camera angle (0-6, Section~\ref{ss:spectra}), and measurement mode (A-D, Section~\ref{ss:spectra}).  For the number of simulations, viewing angles, and observing modes we wish to test in this work, we must measure in total a few times $10^6$ individual lines, so we require a robust, relatively quick way to automate these measurements.  We describe our method in this section.  

Like real spectra, we do not know ahead of time the continuum level $C_{\lambda}$.  So for each spectrum, we must estimate $C_{\lambda}$ on the wavelength range of interest.  This computation is a source of uncertainty for any equivalent width measurement and is especially problematic for the [O II] and H$\delta$ lines because they sit in a portion of the spectrum which varies rapidly with wavelength, especially when Balmer-series absorption is prevalent.  This makes any general continuum-fitting procedure difficult.  

In an attempt to minimize this issue, we follow the approach of many catalogs and calculate $C_{\lambda}$ using an iterative $2\sigma$ rejection filter with a window of width $200$\AA.  Note that these continuum fitting parameters are somewhat arbitrary; we chose them so that the automatic measurement best resembles continua one might draw by hand near all our lines at all times during the simulation.  It should be noted that this procedure somewhat underpredicts the true continuum level, since it is calculating an average of the flux near the lines while the true continuum level is unknown.  Thus our absorption equivalent widths are slightly smaller than a more robust calculation might find.  There are many choices of parameters for this procedure that give reasonable but different results for the continuum estimation, and we find that this introduces uncertainty in a relatively balanced way among the lines of interest near the $4000$\AA\ break, at the $\approx 10\%$ level.  

We then convolve (smooth) the continuum with a normalized $100$\AA \ box and subtract it from the spectrum and convolve (smooth) the result with a normalized $15$\AA \ box to produce the input $S_{\lambda}$ for the line measurement procedure.  For each line, we approximate Equation~(\ref{eq:ew}) by fitting  to $S_{\lambda}$ a Gaussian curve with parameters $A_{\lambda}$, $\sigma$, and $\lambda_{\rm center}$ as the line amplitude, Gaussian width, and center, respectively.   The equivalent width for the fitted curve is then  
\begin{equation} \label{eq:gaussew}
W_{\rm fit} \approx \frac{A_{\lambda} \sigma \sqrt{2\pi}}{C_{\lambda} (\lambda_{center})}.
\end{equation}

This fitting procedure has several advantages over a direct application of Equation~(\ref{eq:ew}).  Most importantly, relative uncertainties to the fit parameters are readily estimated, allowing us to let the fit proceed with minimal preconditions.  We simply throw out line measurements for which the line center is not sufficiently close to the catalog line center, the computed $W_{\rm fit}$ uncertainty is very large, or the Gaussian fit parameters are very highly correlated.  Since we value speed for performing so many calculations, we apply a very simple MCMC procedure from \citet{press}.  We find that a 1000-step burn-in followed by a 5000-step chain more than suffices to estimate the parameters and their relative uncertainties for lines that exist.  

The pipeline was tested extensively by visually comparing the spectral lines to the automatically-calculated EW values.  For lines that are not visually apparent, the parameter uncertainties and/or correlations are tens of percent, and the center of the fitted Gaussian often does not match the line catalog, so we reject measurements that have these properties.  Lines that are obviously significant have small residuals ($\lesssim 5\%$), and nearby/slightly overlapping lines can be separated effectively.  This procedure takes approximately 1 second per spectral line on modern processors: this is adequately fast for our purposes.  Although the uncertainties we measure are not real in the observational sense because our models do not produce measurement uncertainties, their relative magnitudes are very effective at quickly estimating the robustness of our line measurements.  We will use $W_{\rm fit}$ as the primary line strength measurement throughout this paper.  

In this work, for simplicity, we do not use H$\alpha$ as a star formation rate indicator.  This line enters our model spectra from all of the model components mentioned in \S~\ref{ss:sunrise}, and the resolution limitations for the emission components mentioned in \S~\ref{ss:spectra} are exacerbated by AGN contributions and the nearby Nitrogen lines.

Future work on these simulations will include the H$\alpha$ line.  It is useful as an additional observational discriminant against dust-obscured star-forming contaminants to the K+A population.  Moreover, a number of different selection criteria have been applied to select poststarburst galaxies in different observational samples.  In principle our models can mimic any of these, but here we do not consider more than simple EW cuts.  This is sufficient to test K+A formation and discuss their properties, but not for complete comparisons to real samples.  A full experiment on the merits and results of different selection methods is beyond the scope of this paper, but will be the subject of future work.  


\subsection{Modeling and Numerical Considerations} \label{ss:systematics}   \label{ss:convergence}

In \S\ref{ss:sunrise} we described our default model and why we believe this choice to be reasonable for the galaxies considered here.  However, we might expect poststarburst selection to depend sensitively on these assumptions in certain cases, so in this section we discuss briefly how our measurements are affected by different modeling assumptions and numerical resolution.  These variations are motivated by the large uncertainties in modeling the dense, star-forming, obscuring ISM.  Such uncertainties arise because many of the processes that drive the evolution of this medium exist on scales below our resolution or are computationally intractable for such kinds of simulations.  Hence we are limited to using sub-grid models with parameter choices constrained to some degree by observations.  

For example, the rapidly-varying physical conditions of late-stage mergers may imply that the appropriate $f_{\rm PDR}$ varies in a complicated way in both time and space.  Emission lines are strongly suppressed from regions with a high PDR fraction that may exist in extremely dense and massive molecular environments, leading to a situation in which the optical spectrum of a powerful starburst may imply instead a starburst relic.  By contrast, emergent emission lines are strongest when the PDR fraction is low, shortening the length of time a given selection criterion will define a K+A galaxy.  

Thus a value of $f_{\rm PDR}$ near 1.0 may be appropriate for specific galaxies, such as very gas-rich interactions at high redshift \citep[e.g.][]{narayanan09_smg}.  We believe it is unrealistic for this work, since we will focus on times of the simulations that are not extreme starbursts.  However, in order to test the sensitivity of K+A selection to high values of $f_{\rm PDR}$, a number of mergers were run with $f_{\rm PDR}=0.9$.  In general, the K+A lifetimes of these simulations were longer by up to a factor of two than those using smaller values, with a second K+A phase occurring immediately after first pericentric passage and prior to final coalescence (unlike the $f_{\rm PDR} = 0.3$ case).  This can be understood because the first passage produces a weak starburst, but star formation continues; the Balmer absorption strength increases in both cases, but with $f_{\rm PDR}=0.9$, the emission lines from the youngest stars are obscured enough to be considered quiescent by our [O II] emission line cut.  We did not find any simulation with either choice of $f_{\rm PDR}$ where we selected as K+A a snapshot during the starburst.  

Another way to deal with obscuration by dense clumps of gas is to turn off the PDR model and instead treat the entire gas mass as existing in the diffuse ISM phase (sometimes referred to as ``multi-phase OFF'').  In this case, {\sc Sunrise} treats the entire mass of each gas particle as having uniform density over an entire resolution element.  We note that this choice does not necessarily maximize the attenuation of young ($< 10$ Myr old) stars versus the multi-phase ON model, though it does systematically increase the attenuation for stars with ages $> 10$ Myr.  Several simulations were undertaken with this choice, with the effect being to attenuate most of the light from young stars throughout the simulation, including A stars.  Although this choice is unlikely to be justified during the relatively quiescent K+A phase of galaxies, this difference represents a significant modeling uncertainty.

These alternative choices may be appropriate for extreme environments.  However, the duration of the K+A phase is commonly of order 100 Myr or more, significantly longer than the duration of the peak of the starburst where the modeling of dust attenutation is most uncertain.  Thus, while our subresolution models are imperfect at following the ISM during these extreme conditions and should be treated as significant uncertainties, this minimal overlap cannot change general conclusions made about the poststarburst properties of these simulations.  However, they may affect inferences made about specific times during the merger (e.g. the peak of the starburst or AGN activity), so more detailed comparisons may further constrain the sub-grid models.  For example, the K+A pair fraction may rely on a combination of $f_{\rm PDR}$ and progenitor bulge properties (see \S\ref{ss:bulges}), or the dynamics of K+A galaxies may depend on the feedback physics during the final merger coalescence.  

Another important component to the model is the assumed ISM pressure support.  Here, we follow (among others) \citet{cox04_thesis}, \citet{sdh05}, \citet{robertson04,robertson06}, and \citet{hopkins10_bhgas} to apply an equation of state leading to a relatively stable ISM suggestive of local disk galaxies.  We do not vary this model here, but recent studies have relaxed this assumption in order to simulate the clumpy, turbulent ISM observed in disk galaxies at high redshift \citep[e.g.][]{elmegreen07}.  The reader is encouraged to see \citet{bournaud10} and \citet*{teyssier10} (and references therein) for results on how this can affect the evolution of the ISM in simulations of turbulent gas-rich disk galaxies and mergers.  

Although we are unable to fully model the clumpy structure of the ISM, the above model variants serve to bracket the plausible extremes of attenuation.  The default set described in \S\ref{ss:sunrise} and used throughout the rest of this paper is observationally well-motivated, but it is useful to consider the effects of these different extreme assumptions.  In summary, if we assume a maximal fraction by which young stars are attenuated by dust in their birth clouds ($f_{\rm PDR}=1$), galaxies with a high star formation rate more commonly satisfy the K+A criterion.  If instead we assume that all of the dust is distributed more diffusely throughout the ISM (``multi-phase OFF''), then there exists enough dust in our merger remnants to mostly obscure the A stars near the center of the remnant for a large fraction of their lifetimes.  While we do not have a single precise appropriate description of the ISM structure, such false-K+As \citep{brown09} or dust-obscured quiescent galaxies are observationally extremely rare at best, so we are assured that our choices are reasonable for the K+A phase of galaxies and should serve to encompass their real behavior.  

In addition to the modeling parameter choices, we must also choose at what level to resolve the underlying SPH and radiative transfer simulations.   \citet{cox06_feedback} studied this for a suite of merger simulations that were identical in spatial resolution and equal to or worse in mass resolution than those considered here, but with a slightly different feedback model.  They showed that the star formation histories of the {\sc Gadget} merger simulations are the same to within 5\% when the particle number is increased by factors of 2, 4, and 10.  This indicates that our SPH particle number is sufficient to resolve the star formation history of the mergers.  Small differences in the simulated spectra are expected when varying the spatial resolution, because the star formation law we apply depends super-linearly on the gas density, and because the exact distribution of stars and gas will vary with resolution. However, the integrated quantities that will most directly affect the K+A features are typically unchanged at the same level as above.  

\begin{figure}
      \epsscale{1.0}
      \plotone{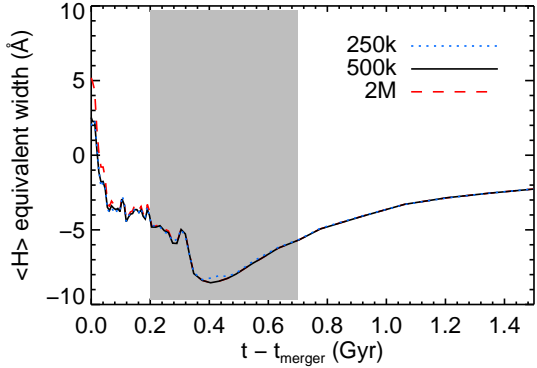} 
      \caption{Here we demonstrate the relative invariance of our lifetime calculations to the resolution of the radiative transfer grid.  We show the same simulation used for row a) of Figure~\ref{fig:bulgecomparison} and row b) of Figure~\ref{fig:cameracomparison} using three different grid refinement settings.  The black curve is the setting used throughout the rest of this paper, and we find that quadrupling the number of cells for a representative snapshot does not dramatically alter the line strength evolution during our times of interest; the gray box is where our fiducial cut might select this simulation as a poststarburst galaxy.   The K+A lifetimes are the same to within a few percent (when all other parameters are fixed).  Note that our definition of equivalent width has absorption as negative values.  \label{fig:rt_resolution} } 
      \end{figure}

The convergence properties of the {\sc Sunrise} radiative transfer approach were demonstrated for the integrated SEDs of isolated galaxies in J10.  In the present work, we use similar grid refinement parameters, such as $\tau_{tol}=1$, and a fiducial resolution setting that leads to $\approx 500$k grid cells for simulations in which the gas fractions are less than or similar to, for example, the Sbc simulation from J10.  During the parts of the merger that K+As could exist in the simulations with our choice of $f_{\rm PDR}$, gas fractions are only a few percent or less and star formation rates are at most a few $M_{\odot}$yr$^{-1}$.  In these cases, diffuse dust attenuation is negligible and the uncertainty in the output spectrum arises principally from the underlying stellar population methods and not the radiative transfer procedures.  To verify similar convergence for our pipeline to produce the post-merger fiber spectra, we re-ran a simulation commonly presented in this work (40\% initial gas, $10^{11} M_{\odot}$ baryonic mass, 1-1 merger) with different resolution settings, giving $\approx 250$k and 2M grid cells for the same simulation snapshot quoted above.  For these three cases, in Figure~\ref{fig:rt_resolution}, we plot the viewing-angle averaged post-merger evolution of the mean EW of the Balmer absorption lines H$\beta$, H$\gamma$, and H$\delta$, which we will use for K+A selection.   The absolute difference in this quantity at our times of interest for K+A selection is less than 12\% between the low resolution and our fiducial setting, and less than 2.5\% between our fiducial and the high resolution setting.  The K+A lifetime we calculate in all three cases is the same to within 0.5\%.  

In addition, the K+A selection criteria among different observations are not identical, and the method we employ here (\S\ref{ss:selection}) is not a perfect match to any one observational scheme.  Unlike many approaches, we do not fit stellar templates to our spectra, and owing to our spectral resolution limitations, we do not attempt to disentangle the absorption component of the Balmer lines from the emission component. Thus our EW measurements can be thought of as essentially the sum of the (signed) emission line flux with the absorption line flux, and differ in this fundamental way from those of, for example, \citet{zabludoff96}, when emission is present.  When the remnant becomes absorption dominated, as for poststarburst galaxies, our models capture the line evolution faithfully.  Future enhancements to our input spectral models will be used to improve upon these measurements.

Another future improvement will be to include in the merger models a realistic treatment of gas recycling during stellar evolution.  In the simulations presented here, a fixed fraction of gas ($0.1$, representing ejecta from Type II supernova) is returned instantaneously to the ISM as part of the stochastic mechanism used to form stars.  However, the complexity and short timescales of gas infall in a merger imply that the evolution of star formation during and immediately after coalescence will be sensitive to the amount, rate, and temperature of gas returned to the ISM from stars formed during the burst.  Moreover, a significant fraction of stellar mass is returned the ISM on long timescales via stellar winds, further motivating the use of accurate gas-recycling methods in studying merger remnants such as K+A galaxies.

\section{Initial Studies} \label{s:results} \label{s:basic_studies}

In this section we apply the above method to a variety of galaxy merger simulations to study their general properties.  Our systematic study of merger parameters follows in \S\ref{s:params}.   

In \S\ref{ss:lightcurves} and \S\ref{ss:lifetimes}, we describe the basic output of our pipeline: the spectral line evolution of each simulation and the ``K+A lifetime''.  We point out in \S\ref{ss:bulges} that stellar bulges stabilize some merging galaxies against K+A creation after their first passage.  In \S\ref{ss:angle}, \S\ref{ss:dust}, and \S\ref{ss:profiles} we employ {\sc Sunrise} to study in detail the effects of viewing angle, dust, and aperture size on K+A selection for several simulations.  The influence of AGN feedback is discussed briefly in \S\ref{ss:agnfeedback}.  

Note that in what follows, we take the convention that equivalent widths in absorption are negative.  


\subsection{K+A Light Curves} \label{ss:lightcurves}

\begin{figure*}
      \plotone{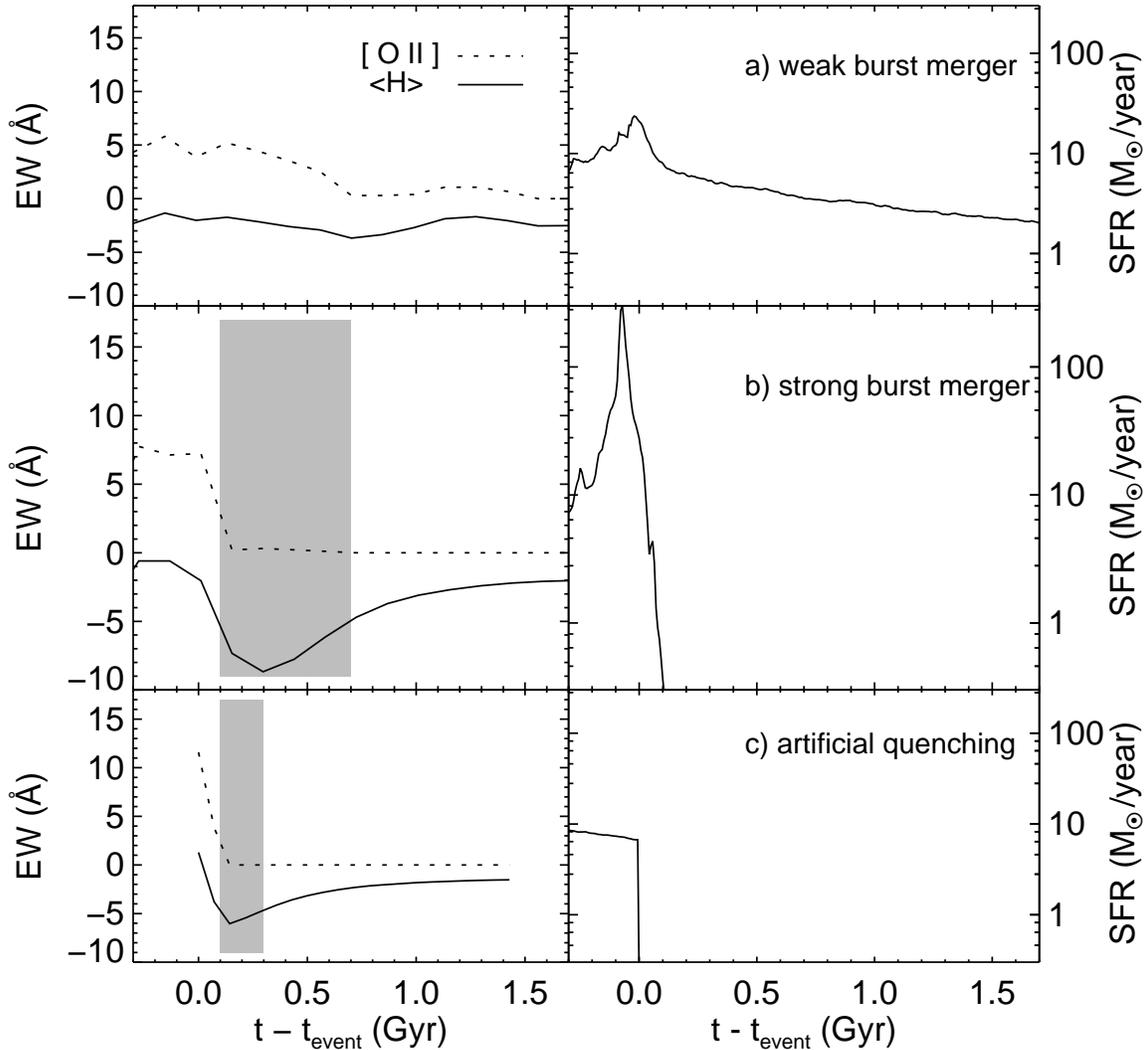} 
      \caption{A summary of different outcomes.  \emph{Left:} The evolution of the K+A parameters as a function of time for three representative simulations.  The dashed curve corresponds to the [O II] equivalent width measurement, and the solid curve indicates the $<$H$>$ measurement, the average of the Balmer lines.  Emission is positive in these figures.  The gray box is where our fiducial cut (-5.5\AA) selects these simulations as poststarburst galaxies.  \emph{Right:} The global star formation rate as a function of time corresponding to the lightcurves in the left column.  \emph{Row (a)} A merger that does not induce a major starburst/shutoff scenario.  \emph{Row (b)} A powerful merger-induced starburst leading to exhaustion of the gas supply.  \emph{Row (c)}  An isolated disk  with $f_{bulge} = 0.25,\ f_{gas} \approx 0.4,\ M_{baryons} \approx 5\times10^{10} M_{\odot}$, for which we have artificially shut off its star formation at $t=0$.  Scenarios (b) and (c) evolve through a K+A phase.  \label{fig:genericlightcurves} } 
      \end{figure*}

The fundamental output of our procedure in \S\ref{s:methods} is a catalog of line strengths at all times during a simulation.  We keep catalogs for 7 viewing angles and for our 4 spectrum models as described in \S\ref{ss:spectra}.  An illustration of this output is shown in Figure~\ref{fig:genericlightcurves}, where we compare the evolution of these line strengths to the global star formation history in the {\sc Gadget} simulation.  We plot the data for a single camera angle and our default Spectrum A.  In general, the star formation histories of our merger simulations range between those that look like row (a) and those that look like row (b).  The only difference between these two simulations is the relative orientation of the merging disks at the start of the simulation.  Thus, as we vary all merger parameters, we expect to see a wide variety in the strength and duration of the K+A spectral features.  

This forms a key result that we will focus on in \S\ref{s:params}: a merger between a given pair of progenitor disks produces a ``maximal'' K+A galaxy only a fraction of the time, because the K+A features require a particular strength and shape to the star formation history.  

For mergers that do not drive a substantial starburst, for example row (a) of Figure~\ref{fig:genericlightcurves}, obvious star formation continues well after the nuclei merge.  In the spectra we measure, the [O II] emission line strength does not drop to zero even after $\sim$ Gyr (and cosmological accretion may drive star formation at near-constant levels), and our average of the three Balmer lines ($\beta,\ \gamma,\ \delta$), $\langle H \rangle$, does not exceed 3\AA\ in absorption.  For mergers that drive a massive burst, e.g. row (b), the gas is efficiently funneled to the center, where it is either consumed into stars or thrown out by violent feedback.  This sequence of events leads to a rapid decline in the [O II] and Balmer emission line strength, with a nearly simultaneous increase in the strength of the Balmer absorption features as the A stars become the luminosity-weighted dominant stellar component.  

For comparison, in row (c), we artificially truncate an isolated star-forming disk (the same disks merged in the first two rows).  As has been pointed out by other authors \citep[e.g.][and references therein]{poggianti99}, this demonstrates that a K+A galaxy as we have defined it is not necessarily a post-starburst system, but can instead be formed by rapidly shutting off the star formation starting from a specific star formation rate of $\approx 10^{-10}$yr$^{-1}$ (a few M$_{\odot}$yr$^{-1}$ for the galaxy we tested).  Other metrics, such as colors, have been used to distinguish the two scenarios \citep{couch87}, but we do not consider photometry in this work.  The reader is encouraged to see \citet{wuyts09}, \citet{scannapieco10}, and J10 for detailed discussions of common observables measured from merger simulations using {\sc Sunrise}.

\subsection{Stellar Bulges} \label{ss:bulges}

An unexpected property of these line catalogs is the strong effect produced by a stellar bulge.  We naively expected that a simulation leading to a major starburst might pass through a single (or no) K+A phase.  However, our simulations of mergers between disks with no stellar bulge component experience two significant phases of enhanced Balmer absorption strength along with reduced emission lines.  The first occurs immediately after the first close passage of the two progenitors, and the second occurs where we expected it: after the two nuclei merge and produce a starburst.  

\citet{mh94a,mh96} described the effect of a stellar bulge stabilizing the center of a gaseous disk against infall induced during close passages.  Since our default measurement mode is to observe the central few kpc of the brightest galaxy in the image, we are particularly sensitive to differences in the star formation history of the central regions of our interacting galaxies.  Without a bulge, a close passage of a gas disk to a major companion is enough to incite a transient bar and subsequent burst-quench cycle in its center, even though the galaxy continues to form stars at $\gtrsim$ 3M$_{\odot}$yr$^{-1}$.  The bulge suppresses these perturbations and preserves more gas for star formation prior to the final burst.  The torques experienced during the final coalescence of a major merger are sufficient to destroy the disk regardless of the bulge.  

\begin{figure*}
      \epsscale{1.0}
      \plotone{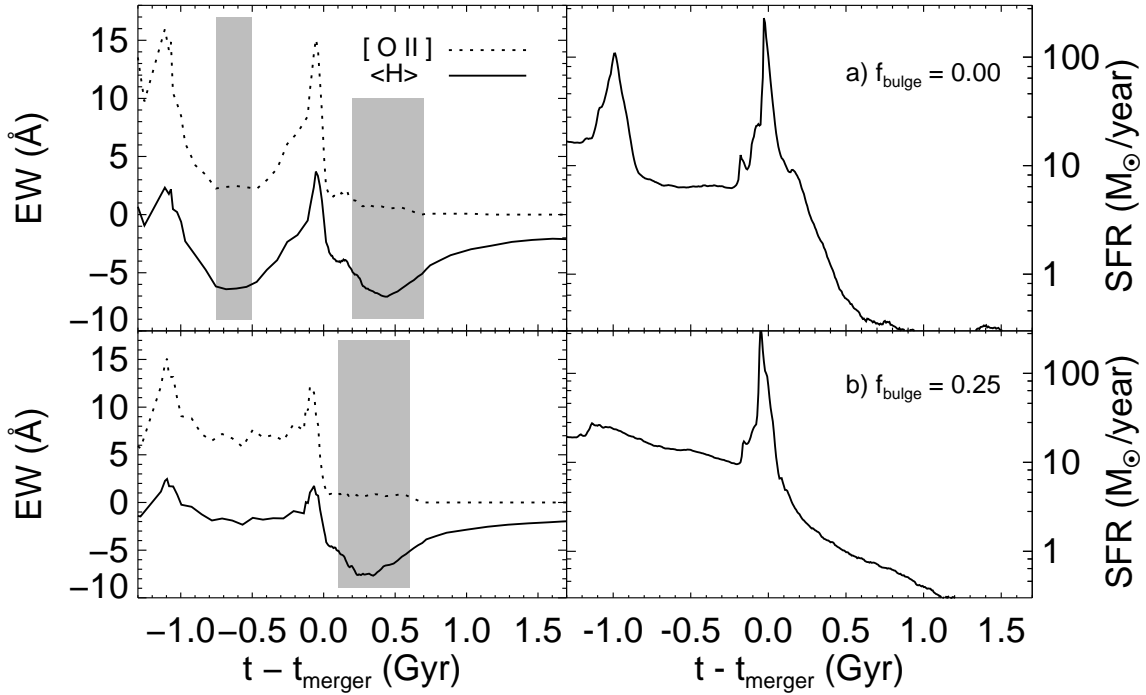} 
      \caption{Demonstration of the effect of bulges on K+A dynamics.  \emph{Left:} The evolution of the K+A parameters as a function of time for two merger simulations identical except for the presence of a stellar bulge.  The gray box is where our fiducial cut might select these simulations as poststarburst galaxies.  The simulation was chosen to be of the ``strong burst" type as shown in Figure~\ref{fig:genericlightcurves}, and we use our default Spectrum A as described in the text.  \emph{Right:}  The global star formation rate as a function of time corresponding to the lightcurves in the left column.  \emph{Top:} $f_{bulge} = 0.0$.  \emph{Bottom:} $f_{bulge} = 0.25$.  The bulgeless merger is a K+A galaxy for more than twice as long as the merger with bulges, inducing a ``binary K+A" phase after first passage, at $t-t_{merger} \approx -0.7$ Gyr.  Despite the global star formation rate being $\approx 6 M_{\odot}$yr$^{-1}$ during this period, the nuclei of the galaxies experience an inflow/burst/quench cycle that happens to dominate the spectrograph aperture.  When bulges are present, this inflow is suppressed, yet the final starburst and K+A phase are roughly identical; see \citet{mh94a,mh96} for a discussion of this phenomenon.  In our simulations, this behavior of a bulge suppressing the first-passage K+A phase while leaving the post-coalescence phase intact is representative of all mergers that experience a significant burst owing to the first passage of the two disks.   \label{fig:bulgecomparison} } 
      \end{figure*}

This effect is shown in Figure~\ref{fig:bulgecomparison}, where we merge two identical galaxies that have a moderate stellar bulge.  In this case, we preserve the total mass (dark matter plus baryons), but add a stellar bulge equal to 25\% of the resulting baryon mass.  The dynamics of the two scenarios (different only in the presence of the bulge component) are similar enough that the nuclear gas infall and starburst at the final merger lead to very similar K+A lightcurves after coalescence.  Thus, the K+A history of a merger between two disks with significant bulges can be approximated by ignoring the pre-merger lightcurves of a simulation between two bulgeless disks.  This is true in all cases where we have mergers both with and without bulges (about 10\% of our sample), and we will make use of this result for the sake of efficiency.  We note that without bulges, the burst induced by this first passage is highly dependent on the choice of initial disk orientation (see \S\ref{ss:orientation}).  

Although K+A galaxies have a slightly higher probability of having a visible companion galaxy \citep[$\sim$ 10\% vs.\ 5\% for non-K+A field galaxies;][]{yamauchi08} than ordinary field galaxies, these bulgeless simulations suggest that if all merger progenitors lack a dynamically important stellar bulge, then the companion fraction for K+As should be of order 20-50\%.  However, a fraction this high is not observed, consistent with the fact that spiral galaxies in the nearby Universe typically contain a bulge.  

This companion fraction may increase at earlier times, where potential progenitors may have smaller bulge fractions.  In addition, this prevalence of a K+A phase for our late-type mergers between first passage and coalescence supports the assertion that some poststarburst+companion systems are dynamically interacting and not all such systems are a superposition of an interloper with a merger remnant.  

\subsection{Selection Criteria and K+A Lifetimes} \label{ss:selection}   \label{ss:lifetimes}

In this paper we adopt K+A selection criteria in the [O II] - $\langle H \rangle$ plane, where $\langle H \rangle$ is defined as the mean equivalent width of H$\beta$, H$\gamma$, and H$\delta$.  This is similar to the cuts used in, for example, \citet{zabludoff96}.  We parametrize our K+A lightcurves by the amount of time that we see a given simulation satisfying a particular set of selection cuts.  For the K+A lifetimes presented in \S\ref{ss:angle} and beyond, we require W$\mathrm{_{[O II]} < 2.0}$\AA\ in emission, and W$_{<H>} < x$\AA\ in absorption, where $x$ is an independent variable.  In what follows we use $-7.5 < x < -3.5$, where negative values signify absorption.  We will refer often to our fiducial cut, where we set $x = -5.5$.


\subsection{Viewing Angle} \label{ss:angle}

A feature of the {\sc Sunrise} method that we employ is the ability to vary the viewing angle of our mock observations.  Any one of our simulations may correspond to different observed galaxies owing to differing three-dimensional spatial distributions of dust, stellar populations, and AGN relative to the viewing angle.  

The placement of our fiber depends solely on the optical flux, so changes in the dust distribution among differing lines of sight will shift the peak of this flux around the image as we vary the viewing angle.  Thus the stellar populations intersected by the aperture vary, and the extent to which we see the galaxy as characteristic of a post-starburst will differ.  In addition, when more gas is present, star formation may continue at higher levels in various parts of the merger remnant.  By looking at images of the merging galaxies, it is clear that in some cases, our aperture intersects the outer regions (for example, a star-forming spiral arm) of one progenitor in front of or behind the center of the one on which we intended to put down the fiber.  In other cases, the final burst simply does not exhaust the gas supply sufficiently rapidly, so star forming clumps continue to rain into the center well after the peak of the burst.  

\begin{figure*}
      \plotone{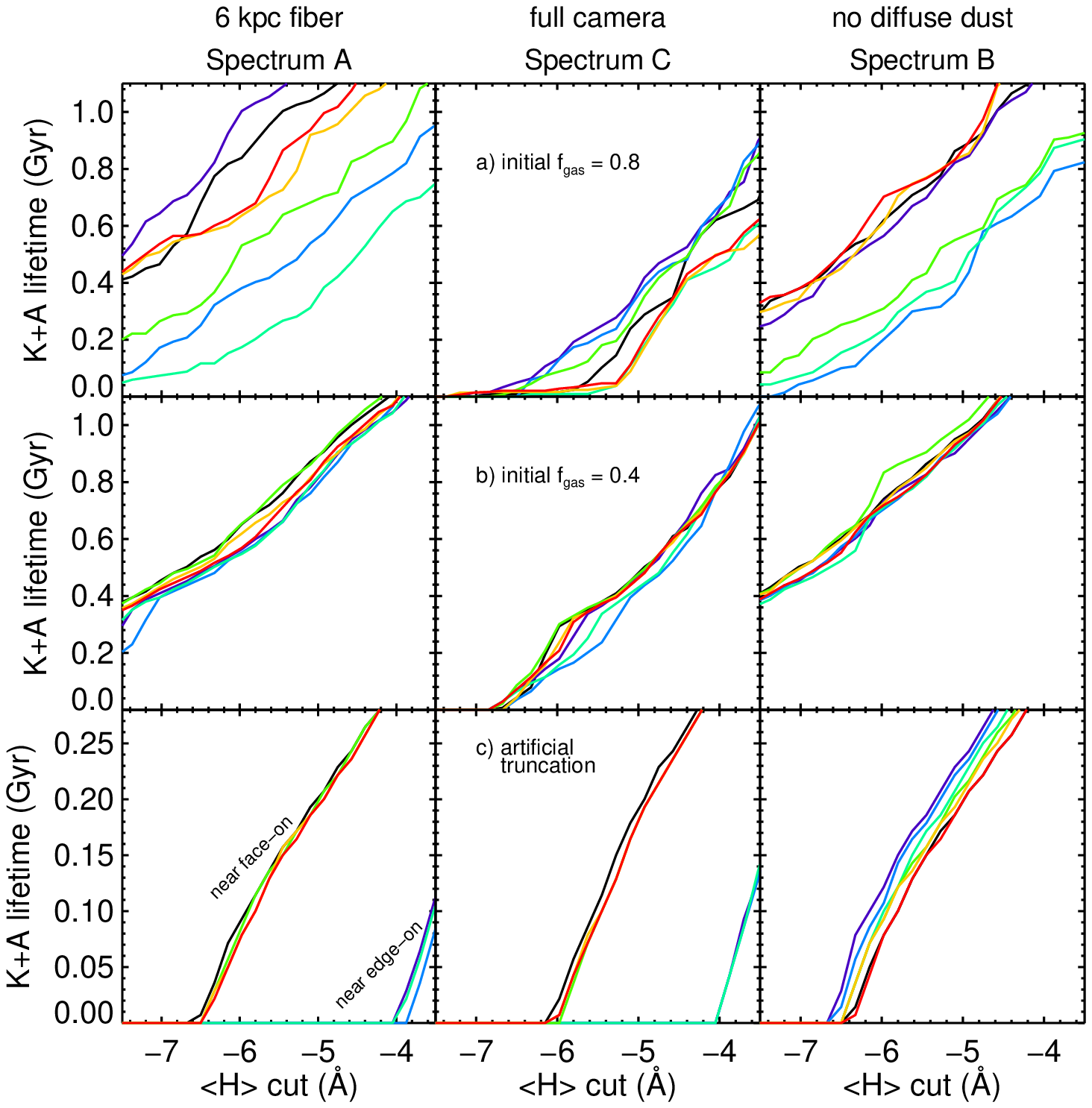}
      \caption{  Demonstration of the effect of viewing angle and dust attenutation on K+A selection, emphasizing the fact that the dust in these remnants is centrally concentrated.  We measure the length of time we view a simulation as a K+A galaxy using our default selection cuts. Each color represents a single viewing angle that is constant across the panels.   \emph{Left:} Our default Spectrum A - measurements made from a spectrum falling within a 3 arcsecond fiber at $z=0.1$.  \emph{Middle:} Our Spectrum C - measurements made with the integrated light of the entire system viewed from different angles.  \emph{Right:} Our Spectrum B - measurements made ignoring the effect of attenuation by dust in the diffuse ISM.  Each row shows one simulation:  (a) gas-rich major merger leading to a strong starburst; (b) identical to (a), except with half the initial gas content; (c) our isolated disk with artificially truncated star formation (the same simulation as row (c) in Figure~\ref{fig:genericlightcurves}).  Thus we can measure the dispersion in K+A incidence with viewing angle for each of our simulations: we keep track of this scatter in Figures~\ref{fig:orientationA}-\ref{fig:bulgesandminors} by plotting its average magnitude as error bars along the top of each panel.  \label{fig:cameracomparison} } 
      \end{figure*}

In the first column of Figure~\ref{fig:cameracomparison}, we quantify this effect on the K+A lifetimes for two representative simulations, as well as our artificially truncated disk for comparison, as viewed through a typical aperture (our default Spectrum A).  Recall that the different spectra A, B, C, and D are listed in Section~\ref{ss:spectra} and are calculated simultaneously for every simulation.  The uncertainty owing to viewing angle correlates most tightly to the gas mass of the disk.  For our major mergers most typical of local spirals (initial $f_{gas} \lesssim\ $40\%, or at the time of merger $ \lesssim $ 15\%), the dispersion owing to angle is of order 10$^8$ years or less.  However, when the initial gas fraction is increased to 80\%, this dispersion becomes of order 10$^9$ years.  In Section~\ref{ss:parameters} and beyond, we will demonstrate this effect for a wider sample of mergers by plotting this as a measurement uncertainty.

We also consider the isolated disk with its star formation artificially cut off to zero when SFR $\approx$ 6M$_{\odot}$yr$^{-1}$: row (c) of Figure~\ref{fig:cameracomparison}.  In this case, the post-starburst component of the galaxy is strongly obscured when looking along the plane of the galaxy, yet easily viewable from higher angles.  Here, dust obscuration prevents the younger A star population from dominating the spectrum.  

In the second column of Figure~\ref{fig:cameracomparison}, we include all the light from the system (Spectrum C) to demonstrate that the magnitude of this dispersion owing to dust obscuration is strongly dependent on aperture size.  This emphasizes the fact that the dust in these remnants is centrally concentrated owing to the global gas infall experienced during the interaction.  We describe the effect of dust in more detail in \S\ref{ss:dust}, and further discuss discuss aperture bias in several simulations in \S\ref{ss:profiles}.

We conclude that when considering merger remnants with small amounts of residual gas, viewing angle effects will not introduce significant errors to the inferred stellar populations.  However, care must be taken when making conclusions about systems that potentially retain a significant gas reservoir.

\subsection{The Effects of Dust Attenuation} \label{ss:modes} \label{ss:dust} 

Here we briefly summarize the effect of dust attenuation on our K+A selection within the framework of our chosen subresolution models.  See \S\ref{ss:sunrise} and \S\ref{ss:systematics} for a more thorough discussion of the underlying ISM modeling and the treatment of dust in dense regions.  Given that we track four separate spectra in each of our simulations (\S\ref{ss:spectra}) in which we vary both the aperture size and presence of dust in the diffuse ISM, it is possible to make approximate conclusions regarding the impact and distribution of dust.  

First off, note that dust influences most aspects of this study.  When significant gas and dust is present, dust along a particular line of sight strongly affects the visible K+A lifetime.  The pre-coalescence binary K+A phase (\S\ref{ss:bulges}) occurs in systems where a gas disk is still intact, so the presence of such a phase is viewer-dependent, and the viewing angle dependence of \S\ref{ss:angle} depends strongly on the gas (hence dust) content and central location.  The visibility and slope of 2-D profiles in the stellar populations (\ref{ss:profiles}) will depend on the dust content and distribution.  We argue in \S\ref{ss:agnfeedback} that any poststarburst dependence on AGN feedback owes primarily to expulsion of the gas and dust from the central few kpc.  

To see this more explicitly, we plot in the rightmost column of Figure~\ref{fig:cameracomparison} the same simulations described in the previous section, except we now ignore attenuation by diffuse dust along each line of sight (Spectrum B).  For simplicity, we focus on this single set of simulations, but the same effect exists (though not shown in this work) for all studies of \S\ref{s:trends}.  In case (a) of Figure~\ref{fig:cameracomparison}, the merger with a high starting gas fraction of 0.8, the K+A lifetimes are in general a few tens of percent longer for each viewing angle when diffuse dust is included (first column) versus ignored (third column).  In these cases dust serves to obscure [O II]-emitting star forming regions that otherwise disqualify the system as a poststarburst, extending the K+A lifetimes.  In addition, we confirm that dust is responsible for the very large line of sight scatter in K+A lifetime seen in Spectrum A for this case.  Note that the two separate groups in Spectrum B result from a particular feature in this merger: the true instrinsic scatters of each group are also much smaller than the first column.  

However, in case (b) where the gas fraction begins at 0.4, the post-burst phases do not coexist with massive star-forming clumps, so the K+A phase cannot be dramatically extended by dust obscuration of these regions as in case (a).  Essentially, the [O II] cut is satisfied with or without the presence of dust.  Instead, the primary observable effect is that the diffuse dust obscures the A star population after the final burst and shrinks the K+A lifetimes by up to $10\%$ for each line of sight.  Note that this effect may also be present in case (a), but is overwhelmed by the effect of star-forming clumps intersecting the aperture.  These tests illustrate that the magnitude and nature of the effect of dust in the diffuse ISM on K+A selection depends in detail on the three-dimensional distribution of the gas and dust with respect to the A star population and star-forming regions.  


\subsection{Radial Profiles and Aperture Bias} \label{ss:profiles}

In \S\ref{ss:angle}, we showed that K+A lifetimes are shorter when a larger extent of the galaxy is subtended by the aperture, because the fraction of the luminosity falling on the aperture owing to A stars, which are more centrally concentrated than the older stars, is reduced.  Thus, generally speaking, all of our simulations are consistent with K+As being the result of a starburst concentrated in the central few kpc.  

\begin{figure}
      \epsscale{1.0}
      \plotone{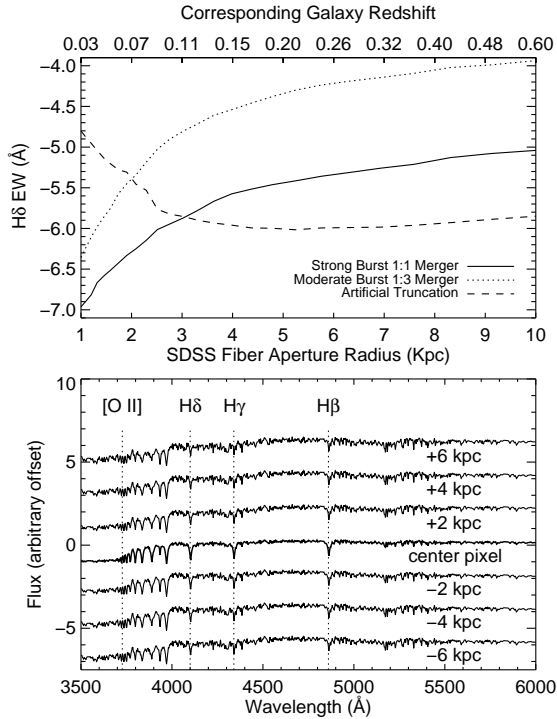}
      \caption{ \emph{Top:} The $H\delta$ equivalent width of single galaxies measured through different spectrograph apertures.  Here we show measurements of two major gas-rich merger remnants (solid and dotted curves), as well as an artificially quenched stable star-forming disk with $f_{bulge}=0.25$ (dashed curve).  Each curve is calculated at a single time in the simulation by measuring the equivalent width of the line using a range of physical aperture sizes (listed in the lower $x$-axis), and averaging over the viewing angles that give a K+A spectrum.   The upper $x$-axis, the galaxy's ``corresponding redshift'', is found by fixing the angular radius of the aperture to 1.5 arcseconds (corresponding to the SDSS fiber), and then computing the redshift for which this angle subtends the physical radius given by the lower $x$-axis.  We note that the two merger remnants have similar trends with aperture size (signal weakens with larger diameter), while the truncated disk has the opposite trend at small radii.  \emph{Bottom: } We demonstrate the ability of {\sc Sunrise} to make integral field spectroscopy maps by plotting the spectrum from seven adjacent pixels in a line near the nucleus of the simulation studied with the solid black curve in the top panel.  We can clearly see the strength of the Balmer absorption lines decreasing and older populations becoming dominant as we move away from the center.  \label{fig:aperture_effect}} 
      \end{figure}

In the top panel of Figure~\ref{fig:aperture_effect} we demonstrate this effect in more detail than is possible with the four types of spectra we will discuss in Section~\ref{ss:params}.  We compute the spectrum for a wide range of spectrograph fiber apertures, and show the spatial variation in the poststarburst spectrum.  The first (top panel) mimics the effect either of using instruments with differing fiber sizes or using the same fiber at a different distance.  The second (bottom panel) mimics the data from multiple pixels available from integral field observations.  The radial dependence of the Balmer absorption is not flat in any of the simulations presented here.  

The curves in the top panel suggest that with a fixed spectrograph aperture, a K+A selection cut applied at $z < 0.1$ will identify a somewhat different class of object than when it is applied at $z=0.5$ or higher.  The difference in H$\delta$ EW between $z=0.1$ and $0.5$ is $\sim 1$\AA\ when measuring a strong-burst merger remnant with the SDSS fiber.  For the merger simulations (neglecting possibly highly obscured contaminants), a constant EW criterion will typically select a more extreme population at higher redshifts.  The effect is opposite for the artificially quenched disk, where an old stellar bulge dominates the spectrum in the very center.  

In the bottom panel of Figure~\ref{fig:aperture_effect}, we demonstrate the ability of {\sc Sunrise} to study the galaxy spectra in a spatially resolved way.  We plot the spectra of seven 2 kpc-sized pixels along a line near the nucleus of one of our poststarburst galaxies (the one from the solid curve, above).  This simulation snapshot is one of our ``strong burst'' mergers and clearly shows a decline in H$\delta$ absorption as we look farther away from the center.  

The gradient is steepest in the central few kpc for this simulated K+A galaxy and many of the ones presented in this paper.  Several observational studies \citep[e.g.][]{goto08, pracy09} have used integral field unit (IFU) spectroscopy to study the radial profiles of the H$\delta$ absorption line, with mixed conclusions regarding the presence of a gradient.  While we have only studied several simulations in this way, our sample is consistent with a clear increase in Balmer absorption EW towards the center of the remnant.  By studying a large number of such simulations, it is possible to estimate what the distribution of gradients is expected to be, and compare in detail with these sophisticated observations, but we leave this endeavor for a future paper.  

\subsection{AGN Feedback}  \label{ss:agnfeedback}

\begin{figure*}
      \epsscale{1.0}
      \plotone{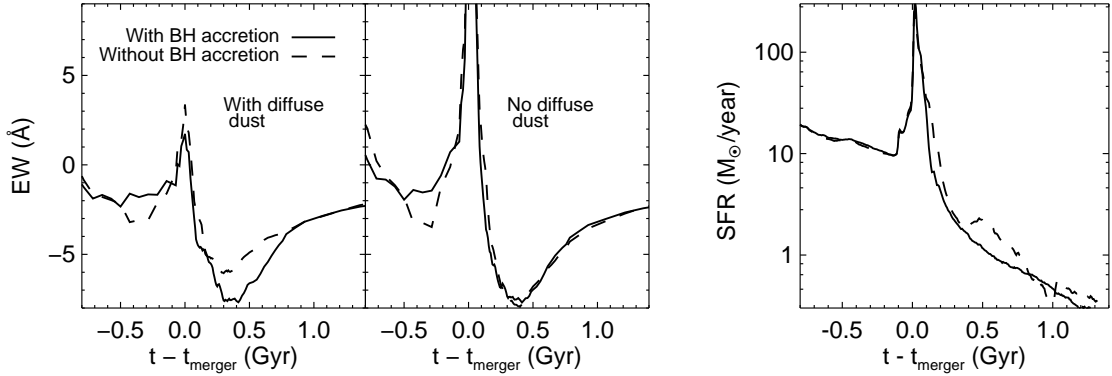}
      \caption{ Here we demonstrate the effect of AGN feedback on the evolution of the Balmer absorption equivalent width.  The solid line represents our default simulation (same as in Figures \ref{fig:bulgecomparison} and \ref{fig:aperture_effect}, 40\% initial gas fraction, 25\% initial bulge fraction, Milky Way- mass, one-to-one merger) with AGN accretion turned ON.  The dashed curve is an identical simulation with AGN accretion turned OFF.  We find that the presence of AGN accretion makes almost no difference to the evolution of the post-merger spectrum when obscuration by diffuse dust is not considered (middle panel).  However, with a realistic treatment of the surviving dust content (left panel), we see that the case with AGN accretion experiences a longer K+A phase with stronger Balmer absorption than the one without this accretion.  We interpret this as AGN feedback dispersing the surviving gas and dust, reducing the optical depth to the young stellar population in the nucleus.  In the right panel we plot the global star formation histories of the merger. }    \label{fig:bh_comparison} 
      \end{figure*}


Several studies, such as \citet{wild09} and \citet{brown09}, have explored the connection between poststarburst galaxies and quasar activity.  While it is thought that AGN-induced outflows carry relatively little gas mass, it is not clear what the effect of this energy deposition will be on the remaining cold gas and subsequent star formation.  Obviously, we have chosen a specific model to approximate these processes, but we can test whether or not it predicts a difference in K+A lifetime from AGN-induced star formation quenching.

Here we briefly demonstrate that in certain cases, the K+A feature depends weakly on the presence or absence of AGN accretion and feedback.  In the example we simulate here, the AGN feedback is responsible for dispersing the remaining gas from the central few kpc, reducing the optical depth to the young A star population, thus strengthening the Balmer absorption in the resulting spectrum.  Dust-removal is the dominant effect: the burst shutdown rate and total mass are not changed enough to affect the K+A lines.  This result is consistent with \citet{wild09} that at least for mergers representative of the low-redshift Universe, the K+A feature is not dramatically affected by AGN feedback.  

We repeat one of our commonly-used one-to-one merger simulations with AGN accretion turned off.  This merger has 40\% initial gas fraction, 25\% bulge fraction, stellar mass of $\approx 4\times 10^{10} M_{\odot}$, and is run on the ``e'' orbit (see below).  In Figure~\ref{fig:bh_comparison} we plot the evolution of the Balmer equivalent width as a function of time through the merger, where the black solid line is our default setting, with AGN accretion turned on, and the black dashed line is with AGN accretion turned off completely.  We show a single viewing angle that is identical in all cases and plot the global star formation rates for comparison.  In the first panel, in which we include the obscuration from diffuse dust, we notice a small ($\sim 1$\AA) difference between the cases with and without AGN feedback, where the merger with feedback experiences stronger Balmer absorption lines.  This suggests that AGN feedback may be contributing substantially to the star formation quenching.

However, in the second panel of the figure, we see that the unobscured stellar populations with or without AGN are nearly indistinguishable after the starburst.  Thus, the slight difference in SFR between the two cases (third panel) is not, by itself, dramatic enough to influence the K+A phase.  Instead of directly shutting down the star formation, the AGN feedback serves to remove obscuring dust from near the young stellar population, enhancing the Balmer absorption features.  We note that this is only a single viewing angle of a single merger, so further study is needed in order to determine whether this effect has reliable observational consequences.  In addition, we leave a systematic study of residual X-ray or LINER emission in simulated K+A galaxies to future work.


\section{Systematic Study of Merger Parameters}   \label{ss:params} \label{ss:parameters}  \label{s:params} \label{ss:trends} \label{s:trends}

In the previous section, we analyzed a number of effects that are relevant for understanding the basic output of our radiative transfer simulations.  In particular, we quantified the effect of stellar bulges, viewing angle, aperture bias, and AGN feedback.  For our fixed choice of physical models, which we argued in \S\ref{ss:systematics} are appropriate for typical local mergers, these effects are dominated by the physical variation in K+A lifetime arising from the variety of late-stage SFHs produced by merging a given pair of disks.  

In this section, we quantify this variation as a function of four parameters that define the initial physical setup of the mergers.  The four parameters we vary are: (a) relative orientation of the disks: all else equal, this parameter determines how effectively the merger will drive gas to the center of the remnant; (b) total mass of the progenitor galaxies; (c) gas fraction at the start of the merger; and (d) the mass ratio between the progenitor galaxies.  A note about gas fractions: real galaxies will accrete matter from the IGM and replenish gas formed into stars, a process we do not include.  Therefore, our initial gas fractions must be set higher than observed gas fractions for typical local galaxies.  The quantity that should be compared to observed gas fractions is the gas fraction at the time of merger; this ranges between 0-30\% for the mergers we consider, consistent with starburst galaxies in the local Universe.  

We point out that we have fixed the orbital parameters of the merger such that the progenitors follow a roughly parabolic orbit with a first pericentric passage at $\approx 7$ kpc.  Changes to this orbit will likely lead to differing K+A lifetimes and line strengths, but for simplicity we limit our study to this initial condition.  

We will assume for the purposes of forming the K+A signatures that the progenitor galaxies have a stellar bulge component that inhibits gas infall owing to a close passage, and consider only the times after final coalescence.  From \citet{yamauchi08} we have seen that this correction can be as high as 5-10\%, but it is within the uncertainties of the merger rates we will be using to estimate the K+A population.  The early-type spiral approximation is reasonable for the majority of mergers in the local Universe, but is likely not correct for mergers at high redshift or specific systems that happen to have low bulge fraction at present times.

\begin{figure}
      \epsscale{1.0}
      \plotone{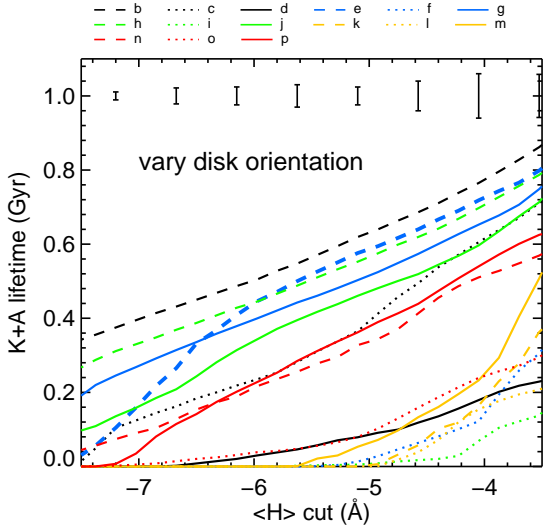}
      \caption{We plot K+A lifetime measurements using Spectrum A (the default) as described in the text, demonstrating the dramatic effect that different orbital configurations have on the duration of the ensuing K+A phase.  Here we vary the initial disk orientation of the galaxies, while keeping fixed the mass of the progenitor galaxies ($M_{baryons} = 4\times10^{10} M_{\odot}$), the gas fraction ($f_{\rm gas, 0} = 0.4$), and the mass ratio (1-1 mergers).  We assume that all progenitor galaxies for major mergers have bulges that inhibit the binary K+A phase prior to final coalescence.  Relaxing this assumption would lead to a set of simulations with an even larger spread in K+A lifetimes owing to the orbit-dependent nature of the burst after first passage.  The bars along the top represent the standard deviation in K+A lifetime (total length $=2\sigma$) owing to viewing angle, averaged over this set of simulations, as shown by Figure~\ref{fig:cameracomparison} for single cases.  Note: these bars do not represent the scatter between the curves plotted here.  \label{fig:orientationA} } 
      \end{figure}

\begin{figure}
      \plotone{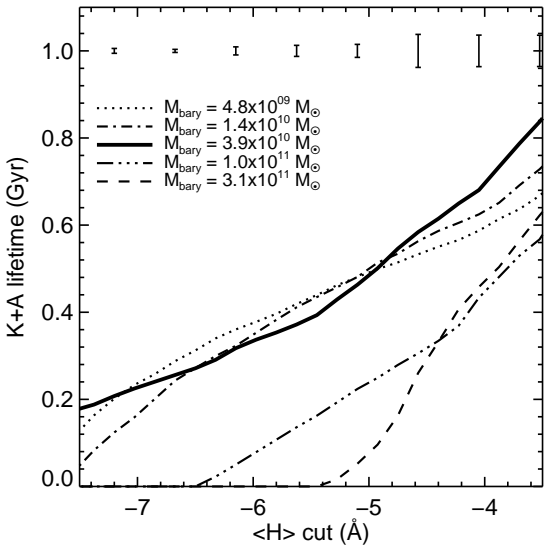}
      \caption{Similar to Figure~\ref{fig:orientationA} except we now fix the orientation (e) and initial gas fraction, but vary the total mass of the merging galaxies.  The bars along the top represent the standard deviation in K+A lifetime (total length $=2\sigma$) owing to viewing angle, averaged over this set of simulations, as shown by Figure~\ref{fig:cameracomparison} for single cases.  Note: these bars do not represent the scatter between the curves plotted here.  \label{fig:massA}} 
      \end{figure}

\begin{figure}
      \plotone{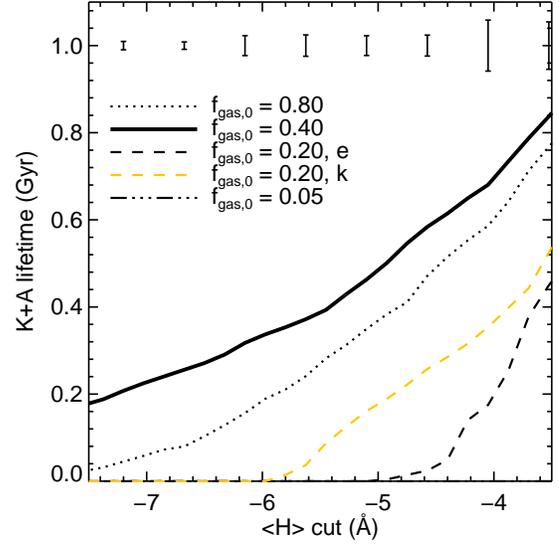}
      \caption{Similar to Figure~\ref{fig:orientationA} except we now fix the orientation (e for all and k for one case) and total mass, but vary the initial gas fractions of the progenitors.  The bars along the top represent the standard deviation in K+A lifetime (total length $=2\sigma$) owing to viewing angle, averaged over this set of simulations, as shown by Figure~\ref{fig:cameracomparison} for single cases.  Note: these bars do not represent the scatter between the curves plotted here. The $f_{gas} = 0.05$ lifetime curve is zero for all values of $\langle H \rangle$ shown here. \label{fig:gasfracA} } 
      \end{figure}

We present the full data from varying the merger parameters separately in the four Figures~\ref{fig:orientationA}-\ref{fig:bulgesandminors}.  In each case, we vary a single one of our four parameters while the other three parameters are held fixed.   When a parameter is held fixed, we use the following values: (a) orientation ``e'' leading to a strong burst; (b) $M_{\rm baryons} = 4\times 10^{10} M_{\odot}$; (c) $f_{\rm gas, 0} = 0.4$; and (d) equal mass mergers.    We summarize these trends and provide fitting functions in Section~\ref{ss:summaries}.  For clarity, we focus only on our default spectrum model (Spectrum A). 

As we have seen, the intrinsic scatter owing to viewing angle may be large: we plot the standard deviation of K+A lifetime owing to viewing angle, averaged over all simulations, as a function of $\langle H\rangle$ at the top of each figure, and find that it can be as high as 100 Myr.  This scatter is a complicated function of both the selection cut and merger scenario.  For mergers that produce shorter-lived K+As ($\lesssim 0.3$ Gyr), the scatter is roughly a constant fraction of the K+A lifetime, and this can be understood as the effect of dust attenuating the lines by a multiplicative factor.  The scatter for these short-lived K+A galaxies tends to be larger than for the long-lived ones because the magnitude of the viewing-angle scatter is directly related to the amount of gas (dust) in the system (Figure~\ref{fig:cameracomparison}), and for cases considered here, mergers that are efficient at consuming gas tend to make longer-lived K+As.  We point out that on average, the viewing-angle scatter is relatively low (less than $\sim$0.05 Gyr in all cases) when considering galaxies selected by our fiducial cut of $\langle H \rangle < -5.5$.

\subsection{Disk Orientation} \label{ss:orientation}

We assume that the relative orientation of disk galaxies in the Universe on large scales is essentially a random variable, as is the viewing angle for which we observe any given galaxy: the merger rates are not fundamentally different for any subset of these tests, and so our simulations here characterize the scatter in K+A lifetime for a given merger event.  However, the merger rate does depend on total mass, mass ratio, and gas fraction (parameters b-c), so ultimately we will use the results from the orientation study to constrain the mean K+A incidence for the remaining parameters.

As shown in Figure~\ref{fig:orientationA}, mergers between otherwise identical disks can produce a very wide range of K+A lifetimes, with a spread of over 500 Myr at our fiducial Balmer cut.  These are the simulations in which we vary only the relative orientation of the progenitor spin vectors.  These orientations are labeled as letters b-p at the top of Figure~\ref{fig:orientationA}, and the exact parameters can be found in Table 1 from \citet{cox06_kinematics}.  Orientations b-h are commonly found in the literature and include special ones like the ``prograde-prograde'' h merger, while orientations i-p are from \citet{barnes92} and chosen so to be unbiased initial conditions.  These vectors are expected to be broadly sampled by real galaxies, and we see from the spread in the figure that both of these orbit subsets span nearly the full range of K+A lifetimes.  We conclude that a given otherwise identical pair of gas-rich progenitors can merge to produce a wide variety of K+A lifetimes.  

In fact, the mean K+A lifetime inferred by this study is $\approx 0.2 \pm 0.05$ Gyr, roughly a factor of 5 smaller than the 1 Gyr typical of canonical estimates!  For mergers of the disks presented in Figure~\ref{fig:orientationA}, our fiducial cuts find K+A lifetimes longer than 0.3 Gyr only about 50\% of the time.  

\begin{figure}
      \plotone{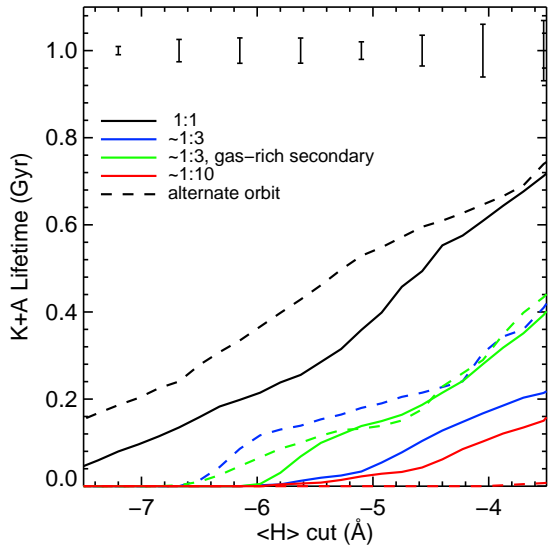}
      \caption{Similar to Figure~\ref{fig:orientationA} except we now fix the orientation (e) and initial gas fraction, but vary the mass of one galaxy to demonstrate the trend of decreasing K+A lifetime with increasing merger mass ratio (\S\ref{ss:modeandminors}).  Solid and dashed curves represent different initial disk orientations that lead to a strong-burst merger when the masses are equal.  \emph{Black - } 1:1 merger, companion $f_{gas, 0} = 0.4$.  \emph{Blue - } 3:1 merger, companion $f_{gas, 0} = 0.4$.  \emph{Green - } 3:1 merger, companion $f_{gas, 0} = 0.8$.  \emph{Red - } 10:1 merger, companion $f_{gas, 0} = 0.4$.  In each case the more massive galaxy has a baryonic mass of $2\times10^{10} M_{\odot}$.  Here we give all the merging galaxies bulges, so the tidal forcing owing to first passage is not sufficient to incite a K+A phase in any of them.  Thus for simplicity we plot the entire evolution of the merger up to $\sim$ 3 Gyr after first pericenter, allowing us to measure the response to multiple passages in the minor mergers.  The bars along the top represent the standard deviation in K+A lifetime (total length $=2\sigma$) owing to viewing angle, averaged over this set of simulations, as shown by Figure~\ref{fig:cameracomparison} for single cases.  Note: these bars do not represent the scatter between the curves plotted here. \label{fig:bulgesandminors} } 
      \end{figure}

In general, the long-lived K+As are ones in which a rapid, strong burst occurred, consuming much of its gas and producing an A-star dominated spectrum.  The short-lived K+As experienced a relatively less violent burst, preserving more gas throughout the merger and creating a weaker signature in the remnant stellar population.  The dust column will be higher in these galaxies and will tend to more strongly obscure the K+A signal.  This gives a shortened observed lifetime and contributes to the separation between classes suggested by Figure~\ref{fig:orientationA}a.   This can also be seen by comparing Figure~\ref{fig:bulgesandminors}a to \ref{fig:bulgesandminors}b.

\subsection{Progenitor Mass} \label{ss:mass}

To isolate the mass dependence, five simulations were run with a single disk orientation (the ``e'' orbit in Figure~\ref{fig:orientationA}), and with a single initial gas fraction (0.4), but with a total mass spanning a factor of $\sim 100$.  One complication is that in real galaxies, properties such as the mass, gas fraction, and bulge fraction are correlated.  However, including all the relevant coupling actually makes the results of such a study harder to interpret.  By isolating these dependencies systematically in this `one at a time' fashion, as we have done here, we hope to gain some physical insight into the real behavior of galaxies.  

In the total baryon mass range $5\times 10^9 - 5\times 10^{10}$ M$_{\odot}$, the strength and duration of the post-merger K+A signature does not vary much (Figure~\ref{fig:massA}), if anything declining slowly with larger mass.   Above roughly $10^{11}$ M$_{\odot}$, the post-merger K+A lifetimes decline rapidly.  This result may seem counter-intuitive because we have fixed the gas fraction and thus expect the burst fractions to be of the same order or larger.  However, the K+A features are sensitive not only to the total burst fraction, but also the detailed shape of the burst star formation history.  The star formation histories of the most massive mergers we simulate are in fact characterized by an extreme burst that may be as large a fraction of total mass as the low-mass cases.  

However, this extreme burst occurs primarily in the mode of a monotonic decrease of global star formation, leading to an efficient, rapid merger.  The final ``starburst'' in these cases is significant, but not distinct enough to form a long-lived K+A galaxy.  By contrast, the less massive mergers experience a burst that better resembles what we might expect for making K+A galaxies: an enhancement followed by a rapid decline in star formation.

We note that the trend of K+A lifetime at high masses does not strongly affect predictions made about the populations observed in large surveys, which will be dominated by significantly more populous lower-mass galaxies.  

\subsection{Gas Fraction} \label{ss:gas}

The trend as a function of gas fraction is shown in Figure~\ref{fig:gasfracA}.  At $f_{gas,0} = 0.8$, the mean K+A lifetime as seen by our aperture is $\sim 100$ Myr shorter than for $f_{gas,0}=0.4$.  However, here we do not simulate many mergers with gas fractions as high as $0.8$, so it is not clear if this trend will actually turn over for the galaxy population as a whole.  We might expect this to happen, as extremely gas-rich merger remants have been shown to reform substantial disks in some cases \citep{robertson06, hopkins_gas09}, and our result is then consistent with the picture that mergers with disky remnants experience shorter and less-robust K+A phases.  As we showed in \S\ref{ss:angle}, a large range of K+A lifetimes will be seen when the gas fraction is this high, with star formation often occurring at high levels after the merger, and high dust column densities obscuring various parts of the burst/post-burst stellar populations.  

When the gas fraction is lower, we recover an expected behavior.  Since the gas reservoir at the time of merger defines an upper limit to the burst mass fraction, for low gas fractions at merger ($\lesssim 0.1$) the resulting starburst will not contribute enough to the spectrum to look like a strong K+A galaxy.

\subsection{Merger Mass Ratio} \label{ss:modeandminors}

Since equal-mass mergers are uncommon, we also consider a number of unequal-mass scenarios.  We define the parameter $\mu$ as the ratio of the baryon mass in the more massive galaxy to the baryon mass in the less massive galaxy.  In Figure~\ref{fig:bulgesandminors}, we plot the K+A lifetimes for a variety of merger mass ratios.  In addition, we consider a second initial disk orientation in order to get a better idea at how this changes the results.  

The K+A lifetime falls off very quickly as we increase $\mu$: on average, three-to-one mergers live as a strong K+A (EW $< -5.5$\AA) for $\lesssim 100$ Myr and ten-to-one mergers $\lesssim 20$ Myr (though note the high variance with viewing angle).  As with equal mass mergers, different orientations introduce a strong dispersion in K+A lifetimes.  

Mergers with a high mass ratio tend to make weaker starbursts \citep[see][]{hernquist89,hm95,hopkins08_minor,cox08_massratio}, owing to the mass-dependence of the response to tidal forcing \citep{ed10}, and this is reflected in the shorter mean K+A lifetimes for these scenarios.  In addition, when a K+A signature is detected for a high-mass-ratio merger, it is often from a single viewing angle (out of seven), and for a single simulation snapshot at a single crossing between the galaxies (out of several).  This leads to rather unpredictable behavior when the interaction is weak, likely in part because the remnants are more disky than the major merger counterparts and hence the burst population experiences greater dust attenuation from certain viewing angles.  Moreover, our method of placing the spectrograph fiber on the peak of the surface brightness leads to noisy measurements as the interactions proceed.  The small bursts in these interactions create stellar populations with complex three-dimensional motions and luminosities that vary rapidly with time.

\section{Summaries and Fitting Formulae}  \label{ss:summaries}

\begin{figure*}
      \epsscale{1.0}
      \plotone{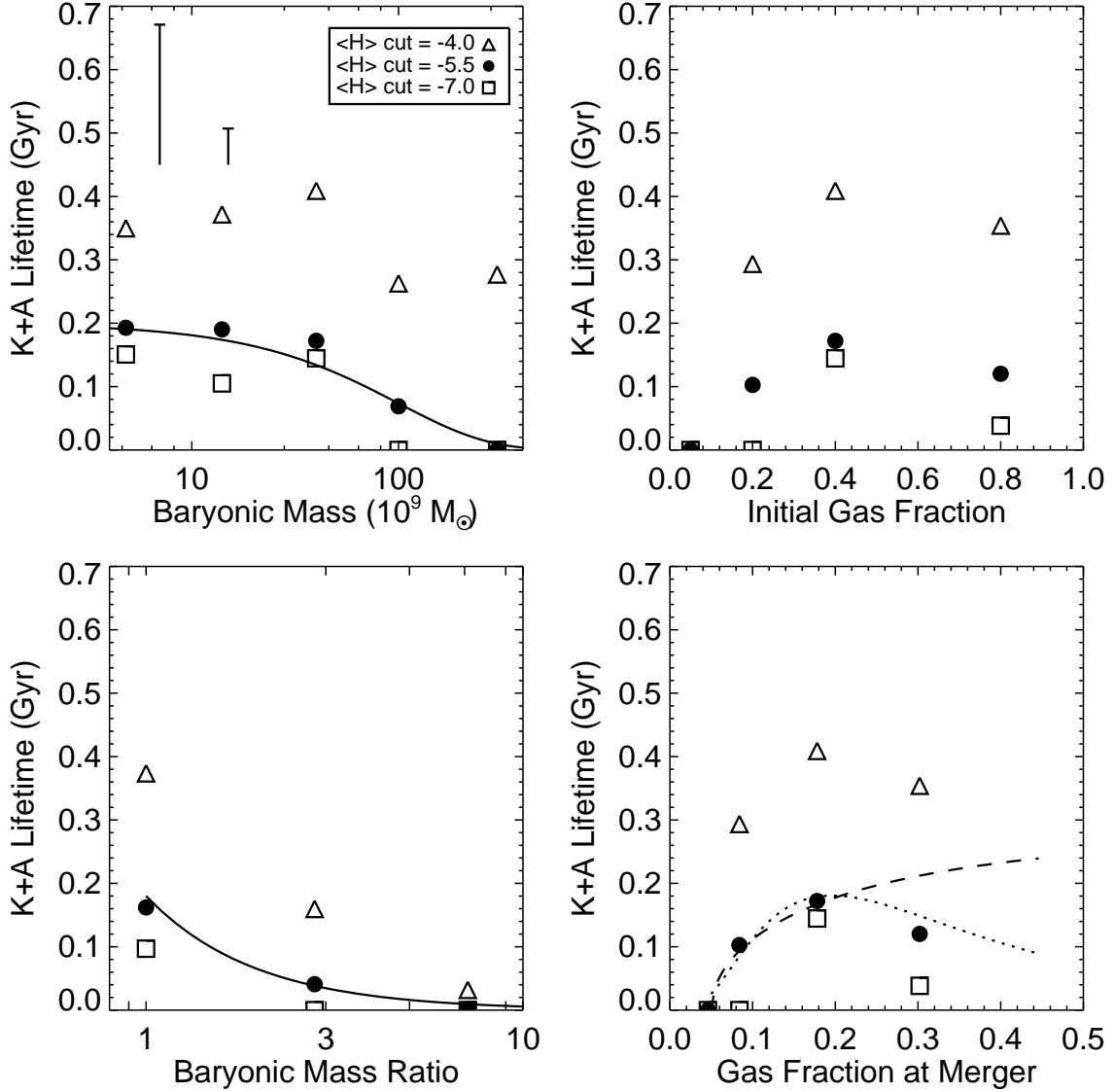}
      \caption{\footnotesize{Summary of K+A lifetime trends with merger mass, gas fraction, and mass ratio.  This strong dependence on progenitor properties drives down the expected per-merger K+A duration, leading to better agreement between K+A abundances and independently-inferred merger rates (Figure \ref{fig:evolution1}).  All points are taken from the data in Figures~\ref{fig:massA}, \ref{fig:gasfracA}, and \ref{fig:bulgesandminors}, and the distribution of lifetimes with disk orientation from Figure~\ref{fig:orientationA} has been used to estimate the trends for the other disk pairs, assuming that the orientation is a uniform random variable selected from the 15 considered in this work.  The filled circles correspond to our fiducial K+A selection cut ($\langle H \rangle \le -5.5$\AA), and the curves, which are described in Section~\ref{ss:summaries}, represent analytic approximations to these points only.  The scatter in lifetime for a specific merger owing to different disk orientations is $\gtrsim 100\%$ (Figure~\ref{fig:orientationA}), completely dominating the scatter owing to other physical effects such as aperture bias and viewing angle.  To show this, we plot a bar representing this dominant intrinsic scatter (total length $=1\sigma$) in the upper left of the first panel.  The smaller bar placed next to it represents the standard error in the estimate of the mean K+A lifetime averaged over the spin orientations obtained by dividing the larger error bar by $\sqrt{15}$.  The size of this scatter is the same to within 30\% for all three selection cuts.  We do not plot the scatter owing to viewing angle.  In the two panels on the right, the four gas fraction cases are the same.  The only difference is that we have shifted the value of $f_{gas}$ to reflect the typical gas fraction at merger for each case (note: the $f_{gas,0}=0.8$ case is sampled most sparsely in our simulations, and so both the K+A lifetime and $f_{gas}(t_{merger})$ are more poorly constrained). }  \label{fig:parametersummaries}}
      \end{figure*}

We show a summary of the trends with the selective parameters mass, gas fraction, and mass ratio, averaged over viewing angle and disk orientation in Figure~\ref{fig:parametersummaries}.  We will focus on our fiducial K+A cut, which is shown as the filled circles.  

One conclusion to draw from this systematic study is that mergers produce a wide variety of K+A behavior, even when the gas content and structure of the progenitor disks are not changed.  The lifetime of this phase is not limited to a fixed single value for all mergers of the same two disks.  In particular, only gas-rich major mergers ($\gtrsim 10\%$ gas fraction at merger) of specific initial disk orientations generate a strong K+A signature ($\gtrsim 5.5$\AA\ EW in Balmer absorption) for a substantial fraction of an A star lifetime.  We find that mergers reproduce canonical estimates of K+A lifetimes of $~1$ Gyr (e.g. from SSPs) only when the dynamics of the merger lead to a burst that is in some sense ``maximal'' for producing the K+A spectral features.  

This means that the dispersion in K+A lifetime for a merger, given its progenitors, is large (of order 100\% in Figure~\ref{fig:orientationA}).  However, when comparing to a large population of galaxies at a snapshot in time, we care mostly about the estimate in the \emph{mean} K+A lifetime for a given pair of progenitors, which is known better by a factor of 3-5 because we can average over these 15 possible spin orientations.  In Figure~\ref{fig:parametersummaries}, a bar representing the $1\sigma$ scatter owing to disk orientation is shown in the upper left corner, and the corresponding (smaller) standard $1\sigma$ error in our estimate of the mean value is shown to the right of it.  

In detail, our study in \S\ref{ss:orientation} finds a mean K+A lifetime that is roughly a factor of five lower than typical estimates of 0.5-1.5 Gyr: here, $T_{K+A} \approx 0.2\pm0.05$ Gyr.  In the next section we show that this rarity of strong poststarburst mergers is consistent to within a factor of a few with nearby large-volume spectroscopic surveys.  

For convenience, we provide example analytic approximations to our general trends that are plotted as the solid black curves in Figure~\ref{fig:parametersummaries}.  

As a function of baryonic mass M above roughly $10^9 M_{\odot}$, we find that the mean K+A lifetimes for our fiducial selection cut ($-5.5$\AA) follow an exponential:
\begin{equation}
T_{M} (M_{baryon}) = 0.2\ \exp \left( - \frac{M_{baryon}}{10^{11} M_{\odot}} \right) \ \rm Gyr.
\end{equation}

As a function of baryonic mass ratio $\mu$ up to at least $\mu \approx 10$, they follow a power law: 
\begin{equation}
T_{\mu} (\mu) = 0.2\ \mu^{-1.5}\ \rm Gyr.
\end{equation}

Most of the simulations we present here have $f_{gas} (t_{merger}) \lesssim 0.35$, so our best-guess values are significantly uncertain beyond this regime.  In the figure, we plot two options, one in which there is a turnover in the trend of K+A lifetimes with higher gas fractions, and one where it is nearly flat at higher gas fractions.   As a function of gas fraction at the time of merger, the former, which provides a better fit to our simulations and is plotted as a dotted line in the bottom right panel of Figure~\ref{fig:parametersummaries}, is approximated by a log-normal curve:
\begin{align}
T_f (f_{gas}) &= 0.034 \log_{10} \mathcal{N} \left ( l_0, \sigma, \right ) \\
&= \frac{0.045}{x} \exp \left (-\frac{\left ( \log_{10}(x) - l_0 \right )^2}{2 \sigma^2} \right ) \rm Gyr,
\end{align}
where $l_0 = - 0.5$ and $\sigma = 0.3$.  The latter option is plotted as a dashed black line in the figure and is approximated by
\begin{equation}
T_f (f_{gas}) = 0.55\ \frac{\sqrt{f_{gas} - 0.05}}{1 + f_{gas}}\ \rm Gyr.
\end{equation}

We derived these trends from a suite of simulations in which we vary only the parameter of interest in the progenitor galaxies.  As we mentioned in \S\ref{ss:mass}, the trend with galaxy mass may be related to the fact that our more massive mergers have differing gas fractions at the time of merger coalescence.  An added complication is that in real galaxies, these two properties are correlated, so we must be careful when combining these trends to ensure that we aren't mis-counting their effects.  Thus, an appropriate combination of these trends may be one in which we take the gas fraction to be the driving parameter over the mass, e.g.:
\begin{equation} \label{eq:combo}
T (f_{gas}, \mu) \approx 0.2\ \rm Gyr \frac{T_f (f_{gas})}{0.2\ \rm Gyr} \frac{T_{\mu} (\mu)}{0.2\ \rm Gyr}.
\end{equation}

In the next section, we use these trends to estimate the K+A fraction given merger rates from an independent source.


\section{Implications for the K+A Fraction} \label{s:counts}
Here we use our lifetime calculations to make a crude estimate of the fraction of all galaxies that are seen as merger-induced K+As.  To calculate this number, we combine our estimated distribution of K+A lifetimes with semi-empirical merger rates.  Thus we must know or estimate the following: (1) The K+A lifetime for every pair of possible interacting galaxies above some mass (flux) limit for a comparison sample, and (2) The merger rate for any given pair of these galaxies at the redshift of interest.  Note that number 2 will change when the prevalent progenitor galaxy properties do, but number 1 is fixed by our calculations for a given pair of galaxies.

\subsection{Merger Rates}
For item (2), the merger rates, we use the semi-empirical models from \citet{hopkins09_mergerbulge,hop10b}.  These models calculate the merger rate per galaxy per Gyr by applying empirical constraints to the halo occupation distribution (HOD) of possible merger progenitors.

We limit our analysis to merger remnants with M$\mathrm{_{baryons}} > 10^{10}$ M$_{\odot}$ and use the median rates quoted in \citet{hopkins09_mergerbulge} for this mass range, because the merger rates are significantly less certain for lower masses and the nearby spectroscopic samples have mass limits of this order.

A crude example of our calculation, undertaken in more detail in what follows, is to multiply a characteristic K+A lifetime by the merger rate.  In the previous section, we found a rough mean lifetime of $\lesssim 0.2$ Gyr for major mergers.  The \citet{hopkins09_mergerbulge} models indicate that the median major ($\mu < 3$) merger rate for galaxies in this mass range is $\rm 0.04\ galaxy^{-1}\ Gyr^{-1}$ at $z=0.1$.  This implies a rough K+A incidence of 
\begin{align*}
&\approx  \rm (0.2\ Gyr) (0.04\ galaxy^{-1} Gyr^{-1}) \\
&= 0.008\ \rm galaxy^{-1}.  
\end{align*}
Note that our K+A lifetime per merger must still be significantly overestimated, because the K+A fraction inferred from large-scale spectroscopic surveys is $\sim 0.001$.  This is because we have not yet taken into account the trends of K+A lifetime with merger parameters, e.g. (most importantly) the merger mass ratio.  Note that the canonical K+A lifetime of $\sim \rm Gyr$, which furthermore does not take into account the dispersion with orbital parameters, would overestimate this fraction even more significantly.

\subsection{K+A Fraction Estimation}

A more accurate calculation must take into account the trends derived in \S\ref{ss:summaries}, because (for example), 3:1 mergers have a lower characteristic K+A lifetime than do 1:1 mergers.  The three merger parameters we consider are the primary ones from \S\ref{s:results}: the total baryonic mass $M_{\rm gal}$ of the merging system, the gas fraction $f_{gas}$ at the time of merger, and the baryonic mass ratio $\mu$.  

We define the K+A lifetime to be the time the galaxy spends with $\langle H \rangle < -5.5$\AA, our fiducial cut, which is similar to the limit imposed in many observational samples.  To estimate (1), ideally we should simulate a complete sample of possible mergers and apply directly our methods to each one.  In \S\ref{ss:params} we performed these calculations for the \emph{axes} of the three-dimensional merger parameter space that we considered, where we take the gas fraction, mass, and mass ratio as the salient properties, plus a range of disk orientations, for a total of $\approx 40$ simulations.  To sample our entire space with a similar quality requires of order $\sim 1000$ more, so in order to reduce the computational expense, we limit ourselves to a rough approximation using the information we already have.  We expect that little more would be gained by calculating the complete sample.

The general trends with each of our merger parameters are readily seen in Figure~\ref{fig:parametersummaries}, so we make the assumption that these trends extend to the unsampled sectors of the parameter space.  For a single simulation in detail, this assumption is obviously incorrect.  To see this, compare the ``k'' orbit from Figure~\ref{fig:orientationA} to the ``e'' orbit for 40\% gas in Figure~\ref{fig:gasfracA} to the ones with 20\% gas in Figure~\ref{fig:gasfracA}.  However, for the parts of our study where we have extended the sample in larger numbers (e.g. Figure~\ref{fig:bulgesandminors}), the trends hold to within the scatter owing to viewing angle.  

An example of this extrapolation was shown in Equation~\ref{eq:combo}.  For the calculation here, we also include a term expressing the mass-dependence.  However, this may not be appropriate in all cases, owing to the previously-mentioned (\S\ref{ss:mass}) coupling between the mass-dependence and the gas fraction-dependence.  Luckily, the steepness of the mass function in comparison samples renders our estimate of the K+A lifetime trend at the highest masses unimportant.  If we assume instead a flat K+A lifetime curve with mass, thus assuming that mass and gas fraction are approximately uncorrelated, our K+A fraction estimates in what follows increase by at most $\sim 10\%$.  

We generate a K+A fraction in the following manner:
\begin{equation}
f_{K+A} = \frac{N_{K+A}}{N_{gal}} = \int p(\mathbf{X})\, T(\mathbf{X})\,  R(\mathbf{X})\, d\mathbf{X},
\end{equation}
where $\mathbf{X}$ represents the full space of merger parameters. $p(\mathbf{X})$ is the fraction of all merger pairs, which we estimate from observations or set by hand, represented by the specific choice of parameters $\mathbf{X}$; this quantity has $\int{p(\mathbf{X}) \, d\mathbf{X} = 1}$, and includes information about the observed merger mass function from \citet{hopkins08}, and the the gas fraction distributions at $z \sim 0.1$ and $1.0$ from \citet{lin08}.  

The quantity $T(\mathbf{X})$ is the K+A lifetime derived from this work for the specific merger $\mathbf{X}$, averaged over the initial disk orientation. Where we have simulations, we use the values from Figure~\ref{fig:parametersummaries}.  Where we do not have direct simulations, we construct $T$ by assuming that the trends in Figure~\ref{fig:parametersummaries} extend trivially to the untested scenarios (e.g. Equation~\ref{eq:combo}).  This introduces uncertainty into our calculation, because we cannot be certain that there is not a dramatic departure from these trends.  

$R(\mathbf{X})$ is the merger rate per galaxy per gigayear from \citet{hopkins09_mergerbulge}, and depends on $\mathbf{X}$ only insofar as we have ignored such dependence in choosing $p(\mathbf{X})$.  We make the approximation that the universal merger rates at a given redshift depend only on $M_{\rm gal}$ and $\mu$, but not separately on the gas fraction.  Since we have chosen to consider a fixed mass range for which the merger rates are compiled, the rates we apply will depend only on $\mu$.

The absolute merger rates for all mergers with $\mu > 10$ are about twice that for ``major'' mergers ($\mu > 3$) alone, but the K+A lifetimes of the major mergers are $\gtrsim 2-5$ times that for minor mergers (\S\ref{ss:modeandminors}) using these definitions.  Thus we might expect that the local K+A population is composed of roughly equal numbers of major and minor merger remnants.  

Within the ``major'' merger range, our lifetime estimates are still rather different between $\mu = 1$ and $\mu = 3$, so we need an estimate of the relative size of each population.  For this purpose, we simply choose two test values by hand to demonstrate the sensitivity of our models: $N_{\mu = 3} = 3 N_{\mu = 1}$ or $ N_{\mu = 3} = 20 N_{\mu = 1}$.  This choice of major merger mass ratio distribution has an effect at a factor of $\sim 2$ level, because this is the corresponding ratio in the K+A lifetime calculations.  

Since our lifetimes also depend strongly on $f_{gas}$, we need to apply an estimate of the gas fraction distribution in our mass range of interest.  To calculate $p(f_{gas})$ as a function of redshift, we assume gas fraction distributions roughly consistent with observed gas-rich/gas-poor merger fractions from \citet{lin08}.  These observations are in good agreement with semi-analytic models, e.g. \citet{kb03}, for halo masses of $\sim 10^{13} M_{\odot}$.

For $z\approx0.1$, where observations suggest that the total mass density in the mass range of interest is split roughly equally between gas-rich disks and gas-poor ellipticals, we assume that merger progenitors are 50\% ellipticals and 50\% spirals, and strongly prefer gas fractions at merger that are $\lesssim 0.1$.  We assume that gas-poor mergers do not form K+A galaxies (this cutoff occurs at roughly $f_{gas} = 0.05$ for our simulations).  At $z\approx 1$ for these masses, the observed gas-free merger fraction is $\sim 10\%$, and gas-rich mergers account for almost $70\%$ of interactions (mixed mergers accounting for the rest).  We assume that typical galaxies at this redshift have gas fractions of 0.2 by putting 50\% of mergers into this bin.

We note that the estimates from \citet{lin08} correspond to a mass range somewhat higher than the ones that dominate our comparison samples, so the values we take of the gas-rich merger fraction are probably somewhat conservative.  In addition, the choices we make of the gas fraction bins in which to put each type of merger are meant to be reasonable, but not necessarily precise.  Changes in the exact details of the gas fraction distribution produce K+A fractions that vary within the uncertainties of other parts of our modeling.  A more thorough calculation can be done using such estimates, but we leave this for a more precise study.  

We interpolate between our two redshift points to plot the K+A fraction as a function of redshift in the solid black line of Figure~\ref{fig:evolution1}.  Such a curve should be thought of as the average merger-induced K+A fraction in all environments at a given redshift.  Note that a predicted merger-induced K+A fraction below the observed one is not necessarily problematic, since there may be additional sources of K+A creation such as ram-pressure stripping or other quenching mechanisms.  

The gray box represents the $~0.5$ dex believable range of our best-guess model.  This quantity comes from uncertainty in the merger rates \citet{hopkins09_mergerbulge} in conjunction with the standard error of the mean K+A lifetimes shown in \S\ref{ss:summaries}.  We plot $0.0 < z < 1.2$, but we do not attempt to account for aperture bias, and as noted above, our accounting for gas fraction evolution is a crude one (and we do not account for things like bulge mass or morphological sub-type).  


\subsection{Comparison to Surveys}

\begin{figure*}
      \epsscale{1.0}
      \plotone{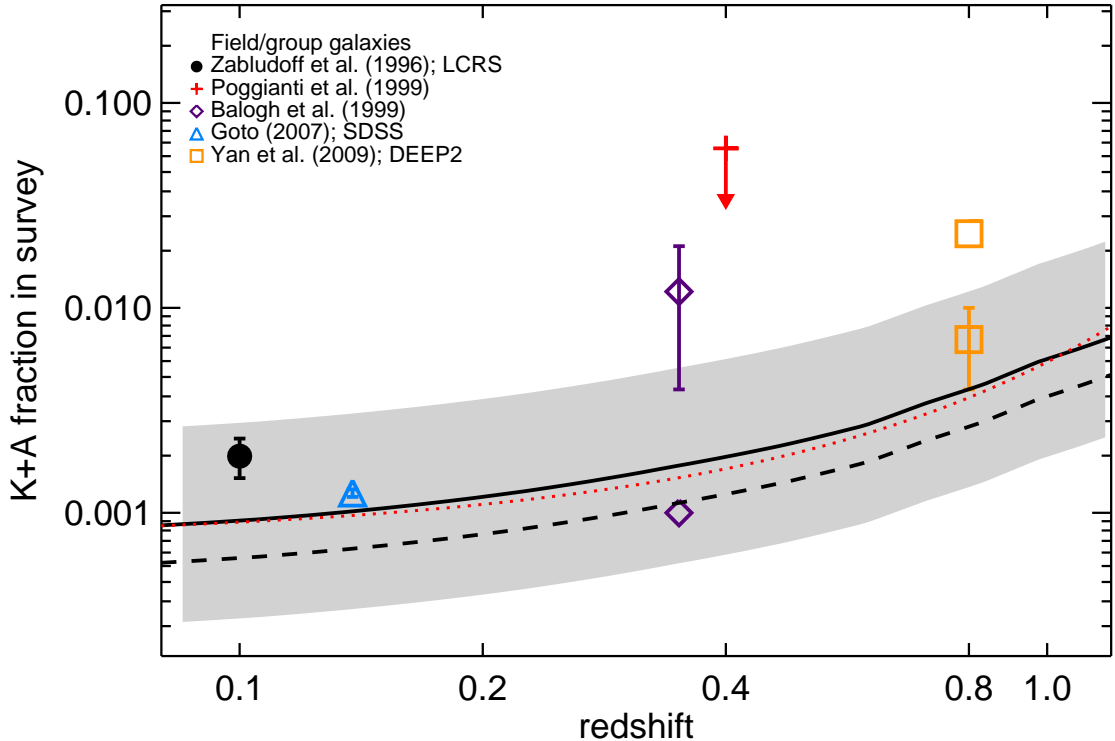} 
      \caption{ \footnotesize{An estimate of the K+A fraction owing to merger-induced star-formation quenching as a function of redshift, for which we find general agreement with similarly-selected samples.  This implies that a typical merger lives as a K+A galaxy for much shorter than the canonical $10^9$ years, bringing the K+A-inferred merger rate up to ones estimated by other means \citep{hopkins09_mergerbulge}.  The black solid line is our estimate when we assume that one-to-one mergers are three times rarer than three-to-one mergers, and is approximated by the relation $f_{\rm K+A} (t) \propto e^{2(z-0.1)}$ (red dotted curve), importantly reflecting a significant increase in K+A abundance as z increases.  The dashed black line is when we set this factor equal to 20; note that this choice only leads to at most a factor of $~2$ difference.  The gray shaded region represents roughly the $1\sigma$ confidence region of the model result, taking into account the uncertainty in the merger rates summed in quadrature with the scatter in the mean K+A lifetime from this work (each a factor of $\sim 2$ for a total $\sim 3$, or 0.5 dex).  The points and error bars reflect survey data and their corresponding measurement uncertainty, with each survey using their own selection criterion.  Error bars for the large local surveys LCRS and SDSS are lower limits to the uncertainty in the fraction computed as $1/\sqrt{N_{K+A}}$, and all other error bars are as supplied by the authors.  All surveys have somewhat different depth limits, corresponding crudely to $10^{10}\ M_{\odot}$ of stellar mass, and this is the cut we use in the merger rate estimates from \citet{hopkins09_mergerbulge}.  Where a survey is given two points, we plot both the ``bright'' and ``faint'' or total sample provided in the referenced paper.  Our selection criteria mimic most closely the LCRS \citep{zabludoff96} and SDSS \citep{goto07} survey points.  A weaker cut may be more appropriate for some of the surveys for which an $H\delta$ cut of $\approx 3$\AA\ was applied; for instance, \citet{poggianti99}, since we include both their ``k+a'' and ``a+k'' classes in the plotted point.  Figure~\ref{fig:parametersummaries} suggests that a factor of two correction to our model curve may be appropriate in this case (this would be insufficient).  However, our models are much more uncertain in this regime, because a much larger class of mergers can cause such a signal, so we have focused primarily on ones that can satisfy our fiducial cut.  \label{fig:evolution1} } } 
      \end{figure*}

Above, we described our method for estimating the K+A fraction at $z = 0.1$ and $z = 1.0$.  At $z \approx 0.1$, our procedure yields a K+A fraction of roughly $10^{-3}$, and the K+A fraction we estimate at $z \approx 1$ is 0.7\%, almost an order of magnitude higher than at $z=0.1$.

For comparison, we plot in Figure~\ref{fig:evolution1} the K+A fraction with associated measurement uncertainty quoted from a number of spectroscopic surveys.  We acknowledge the complications of this approach as pointed out  by \citet{yan08} and other authors, especially the fact that each survey uses different selection criteria, has different flux limits, and is affected in different ways by cosmic variance and environmental effects.  For instance, the \citet{poggianti99} sample has a less stringent EW requirement than we consider here.  Combining our models with a detailed cosmological framework, it is possible in principle to account for such effects, but we defer this more sophisticated study to future work.  However, even with this basic approach we are able to reproduce some expected results as a rough check of our approach.  

At $z\approx0.1$, our models imply that a large fraction of the K+A population in large-sky surveys such as the Las Campa\~{n}as Redshift Survey \citep[LCRS,][]{shectman96, zabludoff96} and SDSS \citep{york00, goto07} can be accounted for by mergers.  This rough correspondence follows directly from one phenomenon: that the K+A lifetimes exhibit strong dependence on the merger scenario.  The population of mergers experiences significant variation in K+A incidence that depends strongly on (at least) orbit, gas fraction, and mass ratio.  As we showed above, removing these dependencies leads to a predicted K+A fraction much higher than observed.

Between $z=0.1-1.0$ in the field, our example calculation demonstrates strong redshift evolution in the K+A fraction.  Our curve falls somewhat below the surveys at $z \gtrsim 0.2$, though still within reason given the uncertainty in the models, and this might be expected if there is important progenitor evolution in mass, typical environment, bulge fraction, or if the evolution in typical gas fraction is more extreme than we assume.  In fact, as redshift increases, we expect that the progenitor galaxies in corresponding surveys will be somewhat higher in mass and later in type than lower-redshift surveys.  Both of these effects could serve to steepen the predicted curve by up to a factor of $~2$ each toward higher redshift.  Another steepening effect may be aperture bias: in Figure~\ref{fig:aperture_effect} we saw that a typical decrease in observed $\langle H \rangle$ signal between $z=0.1$ and $z\sim1$ (at fixed aperture size) is $\approx 1$\AA.  The appropriate comparison is between the filled circles in Figure~\ref{fig:parametersummaries} and the squares (a stronger cut, which is required at higher redshift to select the same galaxies).  We see that this could be an effect at the level of a few tens of percent.  In addition, the other mechanisms for creating a K+A phase may account for some of this difference.

\section{Discussion} \label{s:discussion} 

It is believed that the K+A criteria select galaxies where star formation has recently and rapidly shut down owing to interactions with other galaxies and/or with a cluster environment.  The probability of becoming a K+A galaxy is increased if the galaxy experiences a period of enhanced star formation prior to the quenching, as opposed to the rapid truncation of a low level of continuous star formation.  Mergers and interactions are a natural way to create these star formation histories.  We have studied this process through a large suite of three-dimensional hydrodynamic simulations coupled with the three-dimensional radiative transfer code {\sc Sunrise}, and we discuss the implications of our results in this section. 

\subsection{Poststarburst Merger Models} \label{ss:counts_discussion}

In combination with \citet{wild09}, this work tested the idea that K+As arise, in many cases, via galaxy-galaxy mergers.  Ultimately, measurements of K+A abundances should be reconciled with realistic models for their formation and event rates.  The observed K+A abundance fraction among different samples is relatively robust insofar as the K+A selection criteria are uniform (though the lack of this uniformity is one of our uncertainties).  However, realistic models for K+A events, specifically their duration, must be used in order to infer any information about the merger or event rate and hence the evolutionary context of these systems.  We have demonstrated an order-of-magnitude calculation that explicitly takes this into account, specifically calculating the K+A incidence with physical merger scenario, and we showed in \S\ref{s:trends} that the lifetime of the K+A phase is a strong function of both merger orbit and progenitor properties.  

In particular, the relative orientation of the spin vectors of the merging disks can be varied to create post-merger star formation histories with a wide variety of Balmer absorption strengths and lifetimes.  This scatter with disk orientation represents a physical scatter in the mean K+A lifetime for a given pair of disks of 0.3 dex.  The K+A lifetime is maximum for equal-mass mergers and for gas fractions at the time of merger between $\sim 0.1-0.2$, but fall of sharply at lower values.  Minor mergers can also lead to a short, viewer-dependent K+A phase (\S\ref{ss:modeandminors}).  While major mergers produce a more robust, longer-lived K+A phase, the relative quantity of minor mergers versus major suggests that the contribution to the merger-driven K+A population may be split roughly equally between major and minor.

Thus, we find that the average maximum K+A lifetime for the merger of a given pair of disks is \emph{at most} $\approx 0.2 \pm 0.05 $ Gyr, a factor of at least several below the canonical lifetime often assumed to be $\sim 1$ Gyr.  This is because not every merger event generates these canonical star formation histories.  By allowing that not every gas-rich merger produces a K+A galaxy of order $1$ Gyr, a fixed observed K+A abundance implies a greater merger rate than a calculation that assumes a lifetime of $\sim 1$ Gyr.  For example, using a 1 Gyr duration, \citet{quintero04} infer a K+A event rate of order $0.01$ per galaxy per gigayear at $z \sim 0.1$, a factor of several below the lower limit merger rate of $\approx 0.04$ per galaxy per gigayear from \citet{hopkins09_mergerbulge}, suggesting that if the latter rate is correct and the typical K+A lifetime is $\sim 1$ Gyr, then there are too few K+A galaxies by at least a factor of a few.  

To reconcile the two, we have estimated the redshift evolution of the merger-induced K+A fraction implied by our simulations in \S\ref{s:counts} and Figure~\ref{fig:evolution1}, using the merger rates of \citet{hopkins09_mergerbulge}.  For simplicity, we considered the strict selection cuts of \citet[e.g.]{zabludoff96}, so that the observed K+A fraction at $z=0.1$ is $\sim 10^{-3}$. The model calculation has a large scatter of $\sim 0.5$ dex, but we find broad agreement with large-area nearby ($z \lesssim 0.2$) spectroscopic samples that use similar selection cuts, suggesting that K+A abundances are consistent with a predominantly merger origin.  

\subsection{Implications for Merger-Induced Evolution}

This result has significant implications for the physical mechanism behind the growth of bulge-dominated galaxies.  For example, this lower characteristic duration per event at fixed abundance implies that a correspondingly larger number of galaxies have passed through this phase.  The mean K+A lifetime of our simulations, weighted approximately to the cosmological progenitor abundances and appropriate for this selection, is $\sim 0.02$ Gyr per event.  This implies that since $z=1$, $\sim 40\%$ of galaxies have experienced a K+A phase, however briefly.  Thus, many merger remnants spend a great deal of time outside of the traditional K+A spectral cuts, a consequence that is not too surprising, given the continuous population of galaxies in the ``poststarburst spur'' in \citet{zabludoff96, yan09} and hints that post-burst populations may exist outside the typical selection regions \citep[e.g.][]{quintero04, yan09}.  

Furthermore, the short-lived merger-induced K+As may be ones that experience less bulge growth than the long-lived ones.  Preliminary evidence suggests that the diskiest merger remnants in our simulations also experience the shortest K+A phases, consistent with the picture put forth by, for example, \citet{barnes92_disks} and \citet{robertson06}, that the merger remnant of gas-dominated disks will have a greater disk component.  Our limited experiments at high gas fraction showed that the K+A phase is shorter when the gas content is large ($\gtrsim 40\%$) at the time of merger, in general agreement with the fact that the star formation halts less suddenly, or not completely, in these systems.  

Hence, further study is warranted to better predict results for or interpret surveys of evolving galaxies at higher redshifts.  Although gas-rich progenitors and mergers are more common at $z > 1$, we have seen that the K+A incidence is a complex function of the gas content, mass, and mass ratio of the merging pair, and the re-formation of gaseous disks will certainly play a larger role during the epoch of peak galaxy assembly at $z \gtrsim 2$.  Thus care must be taken to correctly interpret poststarburst abundance measurements in upcoming higher-redshift surveys.  

\subsection{Resolving the Geometry and Dust} \label{ss:dust_discussion}

Our combined hydrodynamics and radiative transfer approach allows us to study in detail the physical evolution of galaxies and their components through the K+A phase.  Roughly temporally coincident with a significant central starburst, SMBH accretion and feedback, and a rapidly-evolving ISM, merger-induced K+A galaxies result from violent events that are more accurately produced by fully three-dimensional physical models.  Using such models, we have begun to study the effects of dust attenuation, spatial bias, and SMBH accretion on poststarburst galaxies in the context of realistic mock observations of physically-motivated simulations.  

These studies fall into roughly two categories: one, understanding systematic observational effects such as aperture bias and viewing angle, and two, varying the physical setup, such as progenitor properties and models used to evolve the structure of the ISM, e.g. forming stars, accreting gas onto the SMBH, and allowing for feedback.  

Since mergers lead to centrally-concentrated starbursts, we have quantified the effect of changes in the spectroscopic fiber aperture between $\approx 1-10$ kpc on several simulated galaxies in \S\ref{ss:profiles}, and shown that a change of order $\sim 1$\AA\ in $\langle H \rangle$ is expected.  This property may distinguish merger-induced K+As from, for example, ones induced by ram-pressure stripping where smaller or inverted gradients may be expected.  

Dust attenuation and viewing angle can introduce large variations in the perceived strength of K+A indicators, especially when the gas content is large.  In particular, any diffuse dust that survives tends to obscure the central young stellar population and weaken the poststarburst signature.  The scatter in the K+A durations induced by this effect is about three times smaller than the one owing to initial disk spin orientation \S\ref{ss:orientation}, and the overall reduction in K+A lifetimes is an effect at the level of a few tens of percent (see Figure \ref{fig:bulgesandminors}).  

The presence of a stellar bulge suppresses the formation of poststarburst signatures caused by relatively weak perturbations such as those induced during the first passage of a merger in certain orbits (\S\ref{ss:bulges}), and this picture is broadly consistent with K+A pair fractions from \citet[e.g.][]{yamauchi08}.  This effect is dominant in a small aperture of $\sim 6$ kpc versus the integrated spectrum.  Combined with knowledge of galaxy structure such as bulge fractions, spatially-resolved observations compared to simulations such as these may provide further constraints on the properties of merger progenitors.  

In some cases, the presence of AGN accretion and subsequent feedback strengthens the K+A signature of the merger remnant \citep[][and \S\ref{ss:agnfeedback}]{wild09}.  The mergers that produce strong starbursts leading to a K+A galaxy are also ones in which much gas is funneled to the center of the coalescing remnant, so merger-induced poststarburst galaxies are likely ones that also experienced significant AGN activity.  For the simulations we consider here, the effect of AGN feedback is limited.  The ``shutdown'' we see is dominated by simple exhaustion of the gas supply and/or feedback related to the starburst itself.  This is consistent with \citet{wild09}, who found that reducing the feedback strength from the central black hole did not substantially alter the poststarburst properties of such mergers.  However, in at least one example, we find that AGN feedback produces a stronger K+A phase ($\sim 1$\AA\ EW enhancement) of longer duration (perhaps up to $\sim 50\%$) than simulations without.  We conclude that the AGN energy disperses dust and gas remaining in the nucleus and assists in uncovering the A-star population.  The systematic comparison of the K+A strength distribution to a large suite of simulations may place constraints on the strength and nature of AGN feedback.  

Several modeling approximations were made in order to calculate the spectrum of simulated galaxies.  These include a subresolution treatment for the H II and PDR regions surrounding young stars as well as for the state of the ISM within each resolution element.  An important free parameter is $f_{\rm PDR}$, which governs the covering fraction or clearing timescale of the molecular birth clouds of young stars.  We chose to use $f_{\rm PDR}=0.3$, corresponding to a clearing timescale of $1-2$ Myr, as a reasonable approximation for local merger remnants, where vast quantities of molecular gas are not present.  We have also chosen to use the multi-phase breakdown of the SPH gas particles to assign dust masses to either the hot phase or cold phase ISM.  We assume that the attenuation of light by the cold phase ISM is captured entirely by the PDR model and so we do not use the cold phase mass to calculate additional dust attenuation.  In \S\ref{ss:systematics}, we discussed several examples where we have varied these assumptions to see their effect, which is substantial in certain cases.

\subsection{Context of Present Simulations}

In work leading up to the present, similar simulations have been used to quantitatively demonstrate that galaxy mergers contribute to the populations of $z=0$ ellipticals \citep{sdh05a,cox06_kinematics,hopkins06_ell,hopkins08_ell}, warm ultra-luminous infrared galaxies (ULIRGs) at $z\approx0-1$ \citep{younger09}, dust-obscured galaxies \citep[DOGs;][]{narayanan09_dog} and submillimeter galaxies \citep{narayanan09_smg, narayanan10_smg, hayward10, hayward11} at $z\gtrsim2$, and quasars at various redshifts \citep{hopkins05_qevol, hop06_quasar, hh06, li07}.  Here, we have shown that K+A galaxies fit within the same framework.  In particular, we have studied the idea that poststarburst galaxies are predominantly remnants of galaxy-galaxy interactions.  Their properties are such that they will likely evolve into passive ellipticals, and these simulations show that mergers of gas-rich disks are likely contributors to the K+A population, supporting the picture that mergers play a significant role in driving late-time galaxy evolution.  Thus, the work here supports the general picture framed by e.g.\ \citet{hop06_quasar,hopkins08,somer08} that these various phenomena are connected in an evolutionary sequence.

Additional evidence for this general scenario can be seen in the form of a fossil record preserved in the structural and kinematic properties of local early type galaxies.  For example, it has recently been shown that the light profiles of gas-rich merger remnants \citep{hopkins08_mr}, ellipticals with central cusps \citep{hopkins09_cusp}, and ellipticals with cores \citep{hopkins09_core} all show evidence of being two-component systems, with an inner relic starburst left over from a gas-rich merger \citep{mh94b}.  These findings can additionally explain the link between the various types of elliptical galaxies \citep{hh10}, the contribution of merger-induced starbursts to the star formation history of the Universe and the infrared background \citep{hopkins10_semi,hh10b,hopkins06_semi}, and the relationships between supermassive black holes and their host galaxies \citep{hopkinsfp_07a,hopkinsfp_07b}.  The present work reinforces the conclusions derived in these studies by making an explicit connection between the immediate descendants of gas-rich mergers and the population of old elliptical galaxies.


\section{Conclusions} \label{s:conclusions}

We calculated rest-frame optical spectral line catalogs of binary galaxy merger remnants using 3-D Monte Carlo dust radiative transfer calculations applied to fully 3-D hydrodynamic simulations.  Using these, we have studied the properties of poststarburst (K+A or E+A) galaxy models in an attempt to understand them in a context of hierarchical galaxy evolution.  We summarize here our key findings:

\begin{enumerate}

\item{We tested the hypothesis that mergers of spirals contribute meaningfully to the population of K+A galaxies.  We find that realistically-simulated major merger remnants commonly satisfy traditional K+A spectral line selection cuts (\S\ref{s:basic_studies}), and quantified the extent to which each of a large suite of merger scenarios satisfies those cuts (\S\ref{s:trends}).}

\item{The K+A lifetime is sensitive to nearly every property of the merging progenitors, including gas fraction, mass ratio, and especially the configuration of their orbit in space (\S\ref{ss:orientation}).  This follows from the fact that merger dynamics exhibit a continuum of starburst properties, from short and violent events to more prolonged bursts.}

\item{Typical durations of the K+A phase as commonly defined are $\lesssim$ 0.1--0.3 Gyr for the merger of equal-mass gas-rich progenitors (\S\ref{ss:summaries}), significantly shorter than the often-assumed value of $\sim 1$ Gyr.}

\item{The shortness of the typical merger-induced K+A phase owes to the fact that not every merger orbit leads to a major, succinct starburst, even when certain configurations of the same progenitors do so.  Thus, the canonical model of a K+A existing for about $10^9$ years is appropriate only for the relatively few cases that the merger dynamics cause a sufficiently significant increase and/or sufficiently rapid cessation of star formation.}

\item{These results reconcile measured K+A abundances with independently-calculated merger rates (\S\ref{s:counts}, \S\ref{ss:counts_discussion}).  By assuming that all gas-rich mergers satisfy the K+A criteria for 1 Gyr, one underpredicts the merger/event rate by up to an order of magnitude.  By accurately accounting for how the K+A lifetime depends on physical merger scenario, as we have done here and for which we provide approximate fitting formulae in \S\ref{ss:summaries}, we find that the inferred merger/event rate is consistent with other estimates.  Furthermore, our model estimates of the K+A abundance as a function of redshift (Figure~\ref{fig:evolution1}) reflect a rapid increase from $z=0$ to $z \sim 1$.}

\item{The subresolution treatment of dust attenuation remains a significant uncertainty during parts of the merger when much gas and dust is present (\S\ref{ss:systematics}).  However, we believe our choices to be reasonable for the vast majority of K+A galaxies simulated here, because most of the time they are relatively dust-poor.  The effect of dust attenuation within our chosen subresolution framework is small and characterized throughout \S\ref{s:basic_studies} and \S\ref{s:trends}, and summarized briefly in Figure~\ref{fig:bulgesandminors} and \S\ref{ss:dust_discussion}.}

\item{The role of AGN feedback in truncating star formation to produce the K+A features in these simulations is minimal.  The energy released by rapid merger-induced accretion onto the SMBH expels the now centrally-concentrated gas and dust from the merger remnant, slightly enhancing the visibility of A stars produced by the starburst, but does not alter the star formation enough to be observable in the unobscured K+A signature.  However, this work focused primarily on local analogues and a single feedback model, warranting further study.}

\end{enumerate}

With our suite of realistic merger simulations, we have gained insight into how merger-induced poststarburst galaxies evolve and depend on their progenitors and physical models.  We have described and executed a method for deriving the true poststarburst lifetimes of merger remnants and reconciled K+A abundances with independently-estimated merger event rates.  Such techniques can valuably constrain the paradigm of merger-induced galaxy evolution by generating observationally-consistent predictions for the properties of evolving galaxies.  For example, semi-analytic modeling, based on approximate fitting formulae such as the ones we have provided here for K+A incidence, could be used to gain insight into related questions.  Future work with similar techniques is warranted to disentangle the underlying causes of galaxy evolution, constrain physical models, and to inform observations and theories of galaxy formation at ever-higher redshifts.  

Specifically, the role of AGN versus star formation and gas consumption in the formation of ellipticals can be probed by varying the physics applied in large suites of merger simulations \citep[e.g.][]{wild09} and comparing them in detail to observations of the structure \citep{yang08}, metal abundance \citep{norton01}, AGN luminosity and SED \citep{brown09}, AGN/starburst ratios \citep{yuan10}, and ISM properties \citep{rich10} of K+A galaxies, LIRGs, ULIRGs, and others.  The complex state of the gas and stars in galaxies, mergers, and their remnants may depend on the precise nature of AGN and stellar feedback mechanisms, the details of gas recycling by stars, and the physics of the ISM, so that such observations and simulations can together provide useful constraints on the underlying physical models.  For comparisons at higher redshift, mergers between more gas-dominated disks with a turbulent ISM structure can be simulated so that we may interpret correctly observations of mergers and their remnants during the epoch of peak galaxy assembly.  

\acknowledgements

GS gratefully acknowledges helpful discussions with Gurtina Besla, Elena D'Onghia, Tomotsugu Goto, Daniel McIntosh, Desika Narayanan, Paul Torrey, Vivienne Wild, Ann Zabludoff, and Dennis Zaritsky at various points of this work.  PJ acknowledges support from the W.M. Keck Foundation.  We thank the anonymous referee for numerous essential suggestions.  The computations in this paper were run on the Odyssey cluster supported by the FAS Research Computing Group at Harvard University.  



\end{document}